\documentclass[reprint,superscriptaddress,aps,prd,nofootinbib]{revtex4-2}

\usepackage{graphicx}
\usepackage{dcolumn}
\usepackage{bm}
\usepackage{amsmath}
\usepackage{graphicx}

\usepackage{amsthm,amssymb,extpfeil}
\usepackage{mathrsfs, subfigure}
\usepackage{color,multirow,hyperref}
\usepackage{verbatim}
\usepackage{booktabs}
\usepackage{diagbox}
\usepackage{float}
\usepackage[normalem]{ulem}

\newcommand \p {\partial}

\allowdisplaybreaks[3]

\def \bal#1\eal  {\begin{align} #1 \end{align}}
\def\({\left(}
\def\){\right)}
\def\[{\left[}
\def\]{\right]}
\def\<{\langle}
\def\>{\rangle}

\newcommand{\eref}[1]{Eq.~(\ref{#1})}
\newcommand{\f}[2]{\frac{#1}{#2}}
\newcommand{\bim} {\begin{itemize}[noitemsep]}

\newcommand{\eim}{\end{itemize}}
\newcommand{\be} {\begin{equation}}
\newcommand{\ee} {\end{equation}}
\newcommand{\bc}{\begin{center}}
\newcommand{\ec}{\end{center}}

\newcommand{\diag}{{\rm diag}}

\newcommand{\mc} {\mathcal}

\newcommand{\si}{{\sigma}}

\newcommand{\oi}{\omega}

\newcommand{\thi}{\theta}

\begin{document}

\hfill {{\footnotesize USTC-ICTS/PCFT-25-51}}

\title{  $Q$-ball superradiance: Analytical approach }

\author{Guo-Dong Zhang}
\email[]{guodongz@mail.ustc.edu.cn}
\affiliation{Interdisciplinary Center for Theoretical Study, University of Science and Technology of China, Hefei, Anhui 230026, China}
\author{Shuang-Yong Zhou}
\email[]{zhoushy@ustc.edu.cn}
\affiliation{Interdisciplinary Center for Theoretical Study, University of Science and Technology of China, Hefei, Anhui 230026, China}
\affiliation{Peng Huanwu Center for Fundamental Theory, Hefei, Anhui 230026, China}
\author{Meng-Fan Zhu}
\email[]{skyer@mail.ustc.edu.cn}
\affiliation{Department of Modern Physics, University of Science and Technology of China, Hefei, Anhui 230026, China}

\date{\today}

\begin{abstract}

It was recently discovered that waves scattering off a $Q$-ball can extract energy from it. We present an analytical treatment of this process by adopting a multi-step function approximation for the background field, which yields perturbative solutions expressed in terms of Bessel functions. For thin-wall $Q$-balls, the amplification factors reduce to simple sinusoidal functions, which explains the multi-peak structure of the spectrum and identifies the physical quantities that determine it. For instance, at high frequencies, the peak spacing is simply the inverse of the $Q$-ball size. The analytical solution further enables us to delineate the full range of possible amplification factors. For general $Q$-balls, this analytical framework also substantially improves the efficiency of evaluating the amplification factors.

\end{abstract}

\maketitle

\tableofcontents

\section{Introduction}
\label{sec:level1}

A non-topological soliton is a stable, localized field configuration that shares the same boundary condition as the true vacuum, with its stability guaranteed by a conserved Noether charge, rather than a topological charge, of the system \cite{Friedberg:1976me,Coleman:1985ki} (see \cite{Zhou:2024mea} for a recent review). A $Q$-ball is a prototypical example, formed by a complex scalar field with nonlinear self-interactions, in which the energy per quantum is lower than that of a free particle \cite{Lee:1991ax,Bowcock:2008dn,Heeck:2023idx}. For a spherically symmetric $Q$-ball, the field value iso-tropically decreases from a non-zero value at the origin to zero at infinity. The overdense region near the origin is identified as the interior, while the region with rapid variation is referred to as the boundary. Depending on whether the boundary is thicker or thinner than the interior, the configuration may be classified as a thick-wall or a thin-wall $Q$-ball. 

Beyond the basic $Q$-balls, several field-theoretical and phenomenological aspects or extensions of these objects have been explored. Spinning $Q$-balls, which carry angular momentum in real space, have been analyzed in various spacetime dimensions \cite{Volkov:2002aj,Astefanesei:2003rw,Kleihaus:2005me,Dias:2011at,Radu:2012yx,Almumin:2023wwi}. Gauged $Q$-balls, arising from couplings to gauge fields \cite{Rosen:1968zwl,Friedberg:1976az,Friedberg:1976ay,Lee:1988ag,Kusenko:1997vi,Benci:2010cs,Gulamov:2013cra,Gulamov:2015fya,Nugaev:2019vru,Kinach:2024onh}, possess an upper bound on their charge due to the repelling forces from gauge interactions, and can in some cases be regarded as superconducting objects. Couplings to fermionic fields have also been considered \cite{Friedberg:1976eg,Cohen:1986ct,Anagnostopoulos:2001dh,Levi:2001aw,Xie:2024mxr}, which may provide an approximate phenomenological description of hadrons \cite{Friedberg:1977xf,Lee:1978yu,Friedberg:1978sc}. Beyond the classical treatment, quantum effects on Q-balls have also been explored \cite{Tranberg:2013cka,Kovtun:2020udn,Xie:2023psz}. When gravitational effects are significant, the counterpart of $Q$-balls is called $Q$-stars or boson stars, which are viable candidates for exotic compact objects \cite{Friedberg:1986tp,Friedberg:1986tq,Lee:1986tr,Lee:1986ts,Jetzer:1991jr,Visinelli:2021uve,Kunz:2021mbm,Shnir:2022lba,Loiko:2022noq,DelGrosso:2023trq,deSa:2024dhj,Kunz:2024uux,Jaramillo:2024cus}. In the presence of both gauge interactions and gravity known as gravastars \cite{Mazur:2001fv,Mazur:2004fk,Ogawa:2023ive,Ogawa:2024joy}, $Q$-ball–like configurations may even resemble non-singular black holes. Composite configurations known as Charge-Swapping $Q$-balls, formed as quasi-bound states of multiple $Q$-balls, have also been shown to exhibit remarkable features and longevity \cite{Copeland:2014qra,Xie:2021glp,Hou:2022jcd}. Thermal effects further enrich their dynamics: Q-balls can undergo evaporation in a thermal bath, and their lifetime is sensitive to finite-temperature corrections \cite{Laine:1998rg,Pearce:2022ovj}. In addition to these theoretical extensions, $Q$-balls play an important role in cosmology. In particular, they appear naturally in supersymmetric extensions of the Standard Model and are deeply connected with the Affleck–Dine baryogenesis mechanism, thereby influencing both baryon asymmetry and dark matter production in the early universe \cite{Kusenko:1997si,Enqvist:1997si,Fujii:2002kr,Enqvist:2003gh,Roszkowski:2006kw,Shoemaker:2009kg,Zhou:2015yfa,Kawasaki:2019ywz,Gouttenoire:2021jhk,Kasai:2022vhq,ElBourakadi:2023pue}.

The concept of superradiance originates from Dicke’s work on radiation enhancement in a coherent medium \cite{Dicke:1954zz}. Zel’dovich later proposed that a rotating cylinder with absorbing boundary could amplify incident waves, thereby introducing the idea of rotational superradiance \cite{Zeld1,Zeld2}. In general, superradiance arises when the rotation of an object couples with incoming radiation, leading to energy extraction from the system. Related phenomena, such as Cherenkov radiation, Mach cones, and the critical velocity of superfluids, can be interpreted as manifestations of superradiance induced inertial motion \cite{Bekenstein:1998nt}. Superradiance has particular significance in black hole physics, where it provides controlled settings to probe novel particle physics and gravitational scenarios\cite{Brito:2015oca,Teukolsky:1974yv,Cardoso:2004hs,Dolan:2007mj,Arvanitaki:2009fg,Bredberg:2009pv,Arvanitaki:2010sy,Pani:2012vp,Witek:2012tr,Brito:2013wya,Brito:2014wla,Berti:2015itd,Marsh:2015xka,East:2017ovw,Baryakhtar:2017ngi,Baumann:2018vus,Zhu:2020tht,Zhang:2020sjh,Stott:2020gjj,Baryakhtar:2020gao,Mehta:2021pwf,Roy:2021uye,Chen:2022nbb,Siemonsen:2022yyf,Guo:2025ids,Zhu:2025enp}. The presence of an event horizon and the specific spacetime geometry of rotating black holes make it the focus of extensive investigations, providing valuable insights into various relativistic astrophysical processes.

Interestingly, the internal rotation of a $Q$-ball can also induce superradiant amplification of energy for scattering waves \cite{Saffin:2022tub,Cardoso:2023dtm}. Superradiant scattering off a $Q$-ball provides a new way to study the properties of $Q$-balls. In this setting, the particle number is conserved, while the ingoing and outgoing states may differ in their energy or other physical quantities, allowing for amplification through scattering. In 3+1 dimensions, spinning $Q$-balls and their perturbations have been analyzed \cite{Zhang:2024ufh}, while general perturbative analysis on top of a 1+1D $Q$-ball was carried out in \cite{Ciurla:2024ksm}. Superradiance of Friedberg–Lee–Sirlin solitons has also been investigated \cite{Azatov:2024npx,Zhang:2025oud}. Furthermore, internal-rotational superradiance of boson stars, including the effects of scalar self-interactions and real-space rotation, has also been studied in detail \cite{Cardoso:2023dtm,Gao:2023gof,Chang:2024xjp}.

In this paper, we investigate $Q$-ball superradiance analytically, complementing previous numerical studies. Since the $Q$-ball configuration is generally highly nonlinear, an analytical approach becomes feasible if the $Q$-ball background is approximated by a piecewise function. For large $Q$-balls, this corresponds to the well-known thin-wall limit, while for generic $Q$-balls, a multi-step function can be utilized. The analytical approach offers several advantages for understanding the nature of $Q$-ball superradiance. For instance, a notable feature in the spectra of superradiant amplification factors is the presence of multiple peaks. The analytical approach now elucidates the mechanisms underlying these peaks, offering insights that were largely absent in earlier numerical analyses. Moreover, once the perturbative scattering equations are solved analytically, the spectra can be evaluated far more efficiently.

This paper is organized as follows. In Section~\ref{sec:level2}, we introduce the $Q$-ball model with a sixth-power effective potential, focusing on the non-spinning case for general dimensions $d\ge2$. The parameter constraints are derived (see Eq.~\eqref{pararange}), and the $Q$-ball profile is approximated using the $(n+1)$-step function. Section~\ref{sec:level3}, we present the perturbation solutions on top of the approximated background field via series expansion, which can be simplified as linear combinations of Bessel functions. In Section~\ref{sec:level4}, we discuss the relation between the amplification factors of the two single ingoing modes and the outgoing particle numbers. Section~\ref{sec:level5} focuses on the specific case $d=2$ in the ideal thin-wall limit $n=1$, constructing the amplification factors from the explicit perturbative scattering solutions. For large frequency $\omega$ and matching point $r_*$, the amplification factors reduce to trigonometric forms, explaining the appearance of additional extrema as $r_*$ increases. Stricter bounds on the amplification factors are also derived by varying the relevant parameters, and the analysis is further extended to the general cases $d>2$ and $n\ge2$. Appendix~\ref{sec:levela} analyzes in detail the relation between the extrema of the amplification factors and those of the outgoing particle numbers, while Appendix~\ref{sec:levelb} discusses the $d=1$ case. Finally, Section~\ref{sec:level6} summarizes our conclusions.

\section{ Background solutions }
\label{sec:level2}

In this section, we will briefly review the basics of $Q$-ball solutions, described by a complex field in $d+1$-dimensional spacetime with a specific type of potentials. We then introduce discretization for the $Q$-ball profiles, which can be solved analytically and will be used to obtain the analytical results for scattering solutions in the next section.

\subsection{Setup}

We consider a complex field with a global $U(1)$ symmetry in $d+1$-dimensional spacetime, with the effective Lagrangian\,\footnote{We use a mostly positive signature for the spacetime metric throughout and the natural units $\hbar=c=1$.} given by
\be
        \widetilde{\mathcal{ L }} = - \widetilde\p^{{\mu}} \widetilde{\Phi}^* \widetilde\p_{{\mu}} \widetilde{\Phi} - V  ,~~~
        V = \widetilde{m}^2 \big| \widetilde{\Phi} \big|^2 - \widetilde\lambda \big| \widetilde{\Phi} \big|^4 + \widetilde{g} \big| \widetilde{\Phi} \big|^6 ,
\ee
where the parameters are chosen such that $\widetilde{\Phi} = 0$ represents the true vacuum. We introduce the following dimensionless variables:
\be
  x_\mu=  \widetilde{m} \widetilde{x}_{\mu},~~  \Phi =\sqrt{\widetilde\lambda} \f{\widetilde{\Phi}}{\widetilde{m}}  , ~~ g =  \widetilde{g} \f{\widetilde{m}^2}{\widetilde\lambda^2} ,  
\ee
which allows us to work with the rescaled Lagrangian,
\be
\label{Lagstarting}
\mathcal{ L } =-\p^{\mu} {\Phi}^* \p_{{\mu}}{\Phi} - V,~~~
V = \left| \Phi \right|^2 - \left| \Phi \right|^4 + g \left| \Phi \right|^6 .
\ee
To ensure that the potential has a single global minimum at $|\Phi| = 0$, we require that $g > 1/4$. The conserved charge associated with the global U(1) symmetry is 
\begin{align}
    Q = i \int {\rm d}^d x \left( \Phi^* \Dot{\Phi} - \Phi \Dot{\Phi}^* \right) , 
\end{align}
where a dot denotes the time derivative $\Dot{\Phi} = {\partial\Phi}/{\partial t}$. The energy-momentum tensor for the complex scalar field has components,
\begin{align}
	T_{\mu\nu} = \p_{\mu} \Phi^* \p_\nu \Phi + \p_{\mu} \Phi \p_\nu \Phi^* + g_{\mu\nu} \mathcal{ L } ,  
\end{align}
where $g_{\mu\nu}$ is the Minkowski metric. The equation of motion for the field takes the form
\begin{align}
\label{eombg}
\Box \Phi = \frac{\p V}{\p |\Phi|^2} \Phi,
\end{align}
where $\Box$ is the Minkowski d'Alembertian.

We will focus on Q-balls without real space rotation, whose minimal-energy ansatz takes the form:
\begin{align}
    \label{ans::1}
\Phi=\Phi_Q(t,r) = f_Q(r) e^{-i\omega_Q t}, 
\end{align}
where $f_Q(r)$ is the radial profile function. Without loss of generality, we focus on the case where $\omega_Q>0$ in this paper. For $\omega_Q<0$, one can perform the transformation $\omega_Q\to-\omega_Q$ and $t\to-t$ to map the negative-frequency scenario back to the positive-frequency one. For a stable and spherically symmetric $Q$-ball to exist, the frequency $\omega_Q$ must be real and lie within the following bounds \cite{Coleman:1985ki},
\begin{align}
    \omega_Q^2 & > \omega_{\rm min}^2  \equiv \min_f \left( \frac{V}{f_Q^2} \right) = 1 - \frac{1}{4g} , \\
    \omega_Q^2 & < \omega_{\rm max}^2  \equiv \frac{1}{2} \frac{{\rm d}^2 V(f_Q)}{{\rm d} f_Q^2} = 1 .
\end{align}
These conditions also hold in the case of $d=1$. It is straightforward to observe that when $g = 1/4$, the minimum and maximum values of the admissible frequency coincide. This is consistent with the previously established condition that $g > 1/4$ is necessary to ensure the existence of a true vacuum at $\Phi = 0$.

Substituting the ansatz in Eq.~\eqref{ans::1} into the EoM Eq.~\eqref{eombg}, we obtain the explicit form of the field equation,
\begin{align}
\label{eom::1}
\bigg( \p_r^2 + \frac{d-1}{r} \p_r + \omega_Q^2 \bigg) f_Q = f_Q - 2f_Q^3 + 3g f_Q^5.
\end{align}
The boundary conditions can be derived from the asymptotic behavior of the solution at $r \to 0$ and $r \to \infty$.
When $r\to0$, we require that $f_Q$ approaches a constant to prevent the divergence of the term $f'_Q/r$, which is equivalent to imposing $f'_Q\to0$, where the prime denotes the derivative with respect to the radial coordinate $r$. When $r\to \infty$, the field approaches the true vacuum with $f_Q\to0$, and the asymptotic behavior is given by an exponentially decaying profile. The explicit form of the boundary conditions is thus given by,
\begin{align}
f_Q \to \left\{\begin{matrix}
f_0  & \text{ for } r\to0, \\
f_\infty \exp(-\sqrt{1-\omega^2} r)/r^{\frac{d-1}{2}}  & \text{ for } r\to\infty,
\end{matrix}\right.
\end{align}
where $f_0$ and $f_\infty$ are constants. Due to the nonlinearity of the differential equation, we can solve it numerically using the relaxation method, with the associated boundary conditions,
\begin{align}
 \left\{\begin{matrix}
f_Q'(r)=0  & \text{ for } r\to0, \\
f_Q' + \left( \frac{d-1}{2 r} + \sqrt{1-\omega_Q^2} \right) f_Q = 0  & \text{ for } r\to\infty.
\end{matrix}\right.
\label{bc1}
\end{align}

The existence of the $Q$-ball solution can be inferred by a well-known mechanical analogy. For this, let us define an effective potential
\begin{align}
V_{\rm eff} (f_Q) = -\frac{1}{2} \left( (1-\omega_Q^2) f_Q^2 - f_Q^4 + g f_Q^6 \right). 
\end{align}
Treating $r$ as a ``time'' variable and $f_Q$ as the position of a unit-mass particle, the field equation in Eq.~\eqref{eom::1} can be viewed as describing particle motion in $V_{\rm eff}$, subject to a friction term $(d-1)f_Q'/r$ that depends on both ``time'' $r$ and the ``velocity" $f_Q'$. Under this analogy, the particle starts at rest from $f_0$ and asymptotically approaches the origin as $r \to \infty$, with the potential providing the initial acceleration. Depending on whether this transition process is rapid or not, we get thin-wall or, more generally, thick-wall profiles.

To better understand the difference between the thin- and thick-wall limits of $Q$-ball configurations, we define $f_{\rm min}$ and $f_{\rm max}$ as the locations of the local minimum and maximum of the effective potential $V_{\rm eff}$, respectively, and denote by $f_z$ the zero point of the effective potential that lies between these two extrema. These characteristic points are given by
\begin{align}
f_{\rm min}^2 & = \frac{1-\sqrt{1-3g(1-\omega_Q^2)}}{3g}, \notag \\
f_{z}^2 & = \frac{1-\sqrt{1-4g(1-\omega_Q^2)}}{2g}, \notag \\
f_{\rm max}^2 & = \frac{1+\sqrt{1-3g(1-\omega_Q^2)}}{3g},
\end{align}
and the initial field amplitude $f_0$ must lie within the range
\begin{align}
    f_z \le f_0 < f_{\max}.
\end{align}
The requirement for these characteristic points to exist further constrains the coupling parameter $g$, leading to the upper bound
\begin{align}
    g < \frac{1}{4(1-\omega_Q^2)}.
\end{align}
For quick reference, let us summarize the constraints for the relevant parameters
\begin{align}
    \omega_Q \in \left( \sqrt{1-\frac{1}{4 g}} , 1\right), & \quad 
    g \in \left( \frac{1}{4} , \frac{1}{4(1-\omega_Q^2)} \right), \notag \\ 
    f_0 & \in [f_z,f_{\max}), \label{pararange}
\end{align}
where we have chosen $\omega_Q>0$ without loss of generality.

\begin{figure}
	\centering
	\includegraphics[height=9.6cm]{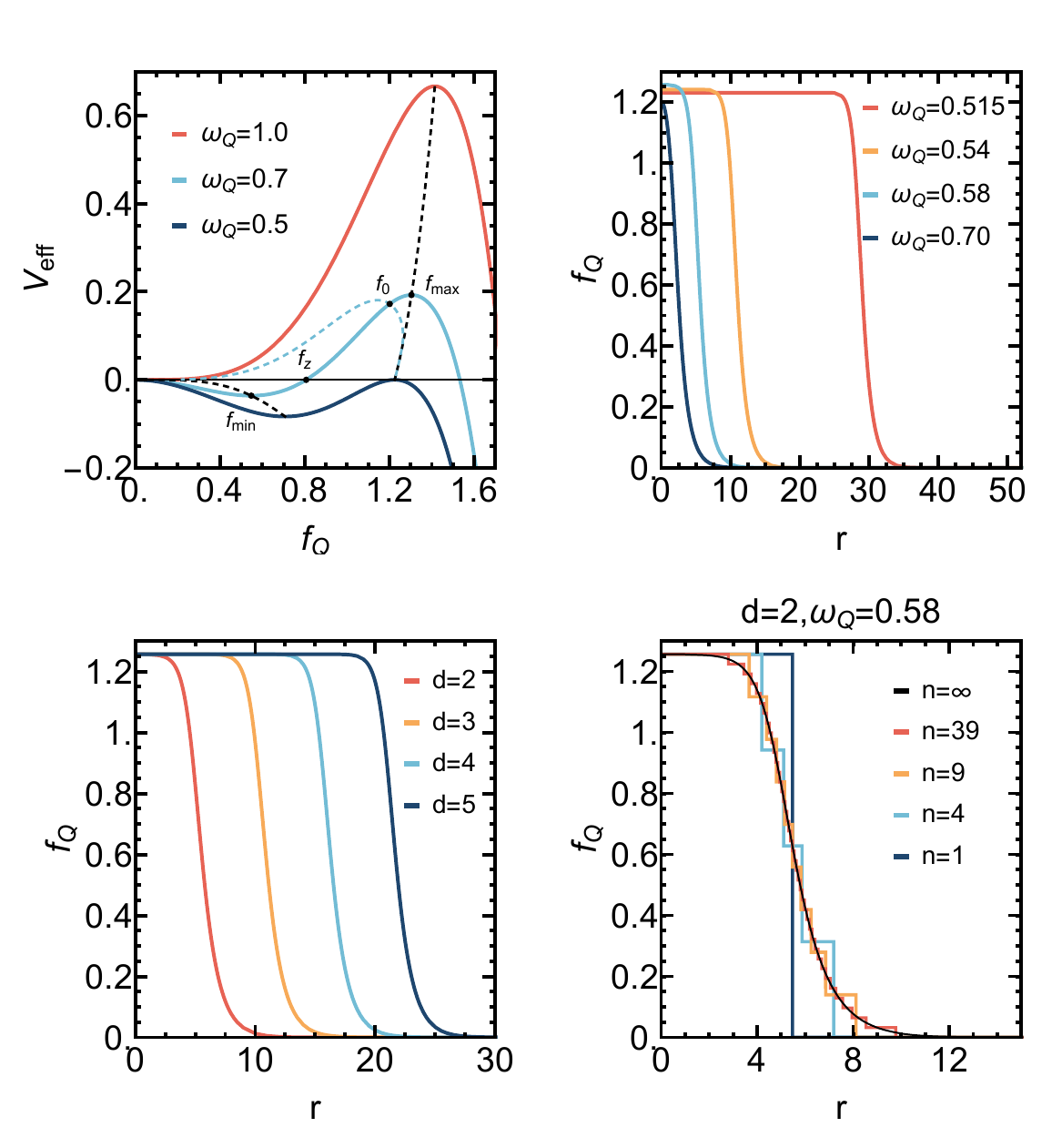}
	\caption{Top left: The effective potential with $g = 1/3$. For varying $\omega_Q$, the potential lies within the region bounded by the red and dark-blue curves. The characteristic points for $\omega_Q = 0.7$ are $(f_{\rm min},f_z,f_0,f_{\rm max}) \approx (0.548,0.807,1.201,1.304)$ with $d=2$. The black dashed lines indicate the minimum and maximum values of $f_Q$ for varying $\omega_Q$, while the light-blue dashed line marks the corresponding $f_0$. Top right: Radial profiles of $Q$-balls for various $\omega_Q$ with $d=2$ and $g=1/3$. Bottom left: Radial profiles of $Q$-balls for various spatial dimensions $d$ with $\omega_Q=0.58$ and $g=1/3$. Bottom right: $(n+1)$-step function approximations of the background field with $g=1/3$, obtained from Eqs.~\eqref{fq::app1} and \eqref{fq::app2}. 
    }
 \label{fig:veff}
\end{figure}

In the top-left panel of Fig.~\ref{fig:veff}, we plot the effective potential, along with the characteristic points $(f_{\rm min},f_z,f_0,f_{\rm max})$. Note that the value of $f_0$ depends on the field frequency $\omega_Q$ and the spatial dimension $d$; here, we choose $d=2$ for illustration. As $\omega_Q \to 1/2$, the separation between $f_0$ and $f_{\rm max}$ becomes smaller. The point $f_0$ corresponds to the initial field value from which a unit-mass “particle” can precisely reach the origin under the combined influence of the effective potential and friction. As seen in the figure, it is evident that the origin corresponds to a local maximum of $V_{\rm eff}$. If the particle is released from $f_0-\epsilon$, with $\epsilon$ a positive value, it lacks sufficient energy to overcome the potential barrier and friction, and eventually settles at $f_{\rm min}$ due to friction. Conversely, if released from $f_0+\epsilon$ (neglecting the change in velocity), it can overshoot the origin and come to rest at $-f_{\rm min}$ due to friction. In general, there also exist special initial values that allow the "particle" to pass through the origin exactly multiple times before finally coming to rest at the origin, corresponding to excited $Q$-ball solutions with multiple nodes. However, the excited Q-ball is an unstable solution with higher energy, and thus will not be considered further in this work \cite{Lee:1991ax,Mai:2012cx,Almumin:2021gax,Zhou:2024mea,Chen:2025oxo}.

In the top-left panel of Fig.~\ref{fig:veff}, a smaller separation between $f_0$ and $f_{\rm max}$ implies a smaller initial “acceleration” for the particle, resulting in a longer time to reach the origin, which corresponds to the thin-wall limit of the $Q$-ball. In contrast, a larger separation leads to a greater initial “acceleration”, and the particle reaches the origin more quickly, corresponding to the thick-wall limit. When $\omega_Q = 0.5$, the effective potential becomes zero at $f_{\rm max}$, and the particle lacks sufficient energy to reach the origin in the presence of friction. This sets the lower bound on the frequency for the existence of $Q$-ball solutions.

The top-right panel of Fig.~\ref{fig:veff} shows radial profiles of $Q$-balls for various $\omega_Q$. As $\omega_Q\to1/2$, the solutions approach the thin-wall limit. The bottom-left panel shows profiles for various spatial dimensions $d$ with fixed $\omega_Q=0.58$ and $g=1/3$; as $d$ increases, the enhanced friction term $(d-1)f_Q’/r$ drives the solutions toward the thin-wall regime.

\subsection{ Analytical background via discretization }

The above discussion establishes the setup of the $Q$-ball field equation along with the corresponding boundary conditions, yielding constraints on the parameters $\omega_Q$, $g$, and $f_0$. The problem can be solved numerically. To facilitate the analysis of perturbations around the background $Q$-ball solutions, it is useful to construct an approximate analytical form. We now turn to constructing a analytical approximation of the $Q$-ball solution via discretization, the simplest of which corresponds to the thin-wall approximation.

In the thin-wall limit, one can model the $Q$-ball solution as a simple step function. For a general thick-wall solution, we can adopt a more refined approximation by representing the $Q$-ball profile as a piecewise $(n+1)$-step function, defined as,
\begin{align}
    f_Q & = 
\left\{\begin{matrix}
 \frac{l}{n} f_0 & \text{ for } r_{n-l} \le r < r_{n-l+1}, \\
 0 & \text{ for } r_{n} \le r,
\end{matrix}\right. 
\label{fq::app1}
\end{align}
where $l = n, n-1, \ldots, 1$, $r_0 = 0$. The initial point is $r_0=0$, and the other points $r_{n-l}$ and $r_{n-l+1}$(for $l\ne n$) are defined by
\begin{align}
f_Q(r_{n-l}) = \frac{l+1/2}{n+1} f_0, \quad f_Q(r_{n-l+1}) = \frac{l-1/2}{n+1} f_0.
\label{fq::app2}
\end{align}
The parameter $n+1$ specifies the total number of steps in the approximation, which reduces to the thin-wall profile when $n=1$.

In the bottom-right panel of Fig.~\ref{fig:veff}, we present the $(n+1)$-step function approximation of the background field. This approximation converges to the exact background field in the limit $n\to \infty$. In the following, we first analyze the simplest $n=1$ case for the perturbative scattering solutions, and then generalize the result to any $n$.

\section{Perturbation solutions}
\label{sec:level3}

Having constructed the $Q$-ball solutions and their approximate analytical forms, we now analyze the scattering of small perturbative waves on the $Q$-ball background. In this section, we derive analytic solutions to the perturbative equations using a series expansion, which can be resummed with Bessel functions. The matching with the scattering asymptotics is deferred to the next section. 

\subsection{ Perturbation equations }
\label{sec:level3::1}

We now consider small perturbations $\phi$ on top of the $Q$-ball background solution $\Phi_Q$ obtained in the last section,
\begin{align}
    \Phi = \Phi_Q + \phi .
\end{align}
The linear perturbations satisfy the following equation of motion,
\begin{align}
\Box \phi & = \left. \frac{\p^2 V}{\p \Phi^* \p \Phi} \right|_{\Phi_Q} \phi + \left. \frac{\p^2 V}{\p (\Phi^{*})^2} \right|_{\Phi_Q} \phi^* , 
\nonumber\\
	& = \left( 1 + U \right) \phi + W e^{-2i\omega_Q t } \phi^* , \label{EOM::ptw}
\end{align}
where $U$ and $W$ are determined by the background $Q$-ball solution,
\begin{align}
\label{bg::U}
	U & = \frac{\p }{\p (f_Q^2)} \left( f_Q^2 \frac{\p V}{\p (f_Q^2)} \right) - 1 = - 4 f_Q^2 + 9 g f_Q^4  , \\
	W & = f_Q^2 \frac{\p^2 V}{(\p (f_Q^2))^2} = -2 f_Q^2 + 6 g f_Q^4 .
 \label{bg::W}
\end{align}
Here, both $U$ and $W$ depend solely on the background $Q$-ball configuration. They approach zero as $r \to \infty$, in accordance with the asymptotic behavior of the $Q$-ball amplitude $f_Q$. 

Applying a Fourier transform to Eq.~\eqref{EOM::ptw}, we find that the perturbative field contains two coupled frequency modes: $\omega_{\pm}=\omega_Q\pm\omega$. Hence, we propose the following ansatz,
\begin{align}
    \phi & = \eta_+(\omega, r)e^{-i\omega_+t}+\eta_-(\omega, r)e^{-i\omega_-t}, \notag \\
    & = \left( \eta_+ e^{-i\omega t}+\eta_-e^{i\omega t} \right) e^{-i \omega_Q t}.
\label{ans::phi}
\end{align} 
Moreover, when we consider the case $\omega<0$, we can redefine $\omega$ as $-\omega$ and exchange signatures $+\leftrightarrow -$ in Eq.~\eqref{ans::phi}. Then it returns to the case $\omega>0$. Therefore, without loss of generality, we can focus on the case $\omega>0$ in the following discussion. Note that the two components $(\eta_+,\eta_-)$ exhibit both a scaling symmetry and a $U(1)$ symmetry, the latter arising from invariance under a time shift. (We have restricted the frequencies to the two values for simplicity, while the generic case is a superposition of the two types of modes.) Substituting the ansatz Eq.~\eqref{ans::phi} into the EoM of the perturbation Eq.~\eqref{EOM::ptw}, we obtain the following coupled equations for the two modes,
\begin{align}
    \left( \p_r^2 + \frac{d-1}{r} \p_r \right) \eta_\pm + (k_\pm^2-U)\eta_\pm-W\eta_\mp^*=0,
    \label{EOM::pte}
\end{align}
where $k_\pm^2=\omega_\pm^2-1$ represents the wave numbers of the perturbations $\eta_{\pm}$, respectively. As we seek a propagating solution, we can impose the following physical condition on the wave numbers/frequencies,
\begin{align}
	\left| \omega_Q \pm \omega \right| > 1 . 
\end{align} 

To construct scattering solutions to Eq.~\eqref{EOM::pte}, we must impose appropriate boundary conditions derived from the asymptotic behavior of the scattering waves. As $r\to0$ and for $d\ge2$, in order to avoid divergence in the term ${\eta_\pm}’/r$, the following regularity condition must be satisfied,
\begin{align}
\p_r{\eta_{\pm}}(\omega_\pm,r \to 0)\to 0.
\label{asymp1}
\end{align}
As $r \to \infty$, the asymptotic form of the equations for scattering waves imposes
\begin{align}
\label{asymp2}
& ~~~~~~~~~~~~~\eta_\pm(\omega,r\rightarrow\infty)\rightarrow \eta_\pm^\infty(\omega,r), \\
& \eta_\pm^\infty(\omega,r) \equiv (k_\pm r)^{-\f{d-1}2} \left( A_\pm e^{ik_\pm r} + B_\pm e^{-ik_\pm r} \right) ,
\label{xi1}
\end{align}
where $A_\pm$ and $B_\pm$ are constants representing the amplitudes of outgoing and ingoing scattering waves, and are related to physical quantities such as particle number, energy, and energy flux. With the appropriate boundary conditions, the perturbative solutions can be obtained numerically. Although for spherically symmetric cases the numerical computations are relatively light, they can become time consuming sometimes for spinning cases or if high accuracy is needed. Among various techniques, the relaxation method provides an efficient and reliable approach for constructing these solutions \cite{Zhang:2024ufh,NRC}. In the following, we will take an analytical approach to solve the perturbative equations based on the discretized backgrounds in the last section.

\subsection{ Series expansion }
\label{sec:level3::2}

Having established the perturbative equations and their corresponding boundary conditions, we now consider a series expansion for the perturbative fields on the background, which enters only through the $U$ and $W$ combinations. Let us work with the $(n+1)$-step function background. Substituting the profile in Eq.~\eqref{fq::app1} into Eqs.~\eqref{bg::U} and \eqref{bg::W}, we obtain piecewise constant background coefficients $U^{(n-k+1)}$ and $W^{(n-k+1)}$ that enter the perturbation equations. These coefficients, distinguished by superscript indices to indicate different spatial regions, are given by,
\begin{align}
    U & = 
\left\{\begin{matrix}
 U^{(n-k+1)} & \text{ for } r_{n-k} \le r < r_{n-k+1}, \\
 0 & \text{ for } r_{n} \le r,
\end{matrix}\right. \notag \\
W & = 
\left\{\begin{matrix}
 W^{(n-k+1)} & \text{ for } r_{n-k} \le r < r_{n-k+1}, \\
 0 & \text{ for } r_{n} \le r,
\end{matrix}\right. 
\end{align}
where $k =n, n - 1,  \ldots, 1$, and the coefficients are defined by 
\begin{align}
    U^{(n-k+1)} & = \left. U \right|_{f_Q \to k f_0/n}, \notag \\
    W^{(n-k+1)} & = \left. W \right|_{f_Q \to k f_0/n}. \label{app::pro}
\end{align}

We use a series expansion to construct perturbative solutions. For $n\ge1$, the background field divides the space into multiple regions: $r \in [0,r_1) \cup [r_1,r_2) \cup \cdots \cup[r_{n},\infty)$. In the first region $r\in [0,r_1)$, we adopt a power series expansion to solve for the two components $(\eta_+, \eta_-^*)$. Again, superscript indices are used to label the respective regions, and the expansion takes the form, 
\begin{align}
\label{pow::sol}
    \! \eta_+^{(1)}=c_0^{(1)}+\sum\limits_{l=2}^{\infty} c_l^{(1)} r^l , ~ 
    (\eta_-^{(1)})^*=d_0^{(1)}+\sum\limits_{l=2}^{\infty} d_l^{(1)} r^l, \! 
\end{align}
where the coefficients $c_l^{(1)}$ and $d_l^{(1)}$ vanish for $l < 0$ and $l = 1$, as required by the boundary condition in Eq.~\eqref{asymp1}. All remaining coefficients are complex constants determined by the perturbation equations. Owing to the scaling and $U(1)$ symmetries of the system, the leading coefficient $c_0^{(1)}$ can be normalized to $1$ without loss of generality. However, in this article, we shall retain it explicitly. Substituting the above series expansion into the perturbative equations~\eqref{EOM::pte}, we obtain the following relations valid in the region $r\in [0,r_1)$,
\begin{align}
\left\{
\begin{aligned}
 0 &= \sum\limits_{l=0}^{\infty} \left[ c_{l+2}^{(1)}(l+2)(l+d) \right. \\
 & \quad \qquad \left. + (k_+^2 - U^{(1)}) c_l^{(1)} - W^{(1)} d_l^{(1)} \right] r^l, \\
 0 &= \sum\limits_{l=0}^{\infty} \left[ d_{l+2}^{(1)}(l+2)(l+d) \right. \\
 & \quad \qquad \left. + (k_-^2 - U^{(1)}) d_l^{(1)} - W^{(1)} c_l^{(1)} \right] r^l.
\end{aligned}
\right.
\label{rela}
\end{align}
Since the equations must hold for all $r \in [0, r_1)$, the coefficients of each power of $r$ must all vanish, leading to the following recurrence relations,
\begin{align}
    c_{l+2}^{(1)} = \frac{W^{(1)} d_l^{(1)} - (k_+^2 - U^{(1)}) c_l^{(1)}}{(l+2)(l+d)} , \notag \\
    d_{l+2}^{(1)} = \frac{W^{(1)} c_l^{(1)} - (k_-^2 - U^{(1)}) d_l^{(1)}}{(l+2)(l+d)} .
    \label{rec::1}
\end{align}
Due to the boundary condition in Eq.~\eqref{asymp1}, we have $c_1^{(1)}=d_1^{(1)}=0$, which implies that all odd-order terms vanish. The remaining even-order terms can be recursively determined in terms of $c_0^{(1)}$ and $d_0^{(1)}$ using the recurrence relations above. Thus, for each choice of $c_0^{(1)}$ and $d_0^{(1)}$, the recurrence yields a valid solution. 

The recurrence relations yield a convergent series, as can be seen by the following estimate. From the recurrence equations, we have
\begin{align}
    \! |c_{l+2}^{(1)}| (l+2) (l+d) \le |W^{(1)}| |d_l^{(1)}| + |(k_+^2-U^{(1)})| |c_l^{(1)}|, \! \notag \\
    \! |d_{l+2}^{(1)}| (l+2) (l+d) \le |W^{(1)}| |c_l^{(1)}| + |(k_-^2-U^{(1)})| |d_l^{(1)}|.\!
\end{align}
Adding the two inequalities gives
\begin{align}
(|c_{l+2}^{(1)}| + |d_{l+2}^{(1)}|)(l+2)(l + d) \leq 2 (|c_l^{(1)}| + |d_l^{(1)}|) G,
\end{align}
where we have defined $G = \max\left( |W^{(1)}|, |k_\pm^2 - U^{(1)}| \right)$. Letting $H_l = |c_l^{(1)}| + |d_l^{(1)}|$, then the inequality becomes
\begin{align}
H_{l+2} \leq \frac{2G}{(l+2)(l + d)} H_l.
\end{align}
Iterating this relation leads to the bound
\begin{align}
H_{l+2} \leq \frac{(2G)^{(l+2)/2}}{(l+2)!!(l + d)!!} H_0 \leq \frac{(2G)^{(l+2)/2}}{(l+2)!} H_0,
\end{align}
where $(l+2)!!$ denotes the double factorial. Since the factorial grows faster than any power function, the higher-order coefficients $H_{l+2}$ rapidly decay, ensuring the convergence of the series. Therefore, in practical applications, the series can be safely truncated at finite order in $r$ without loss of accuracy.

It is instructive to first look at the case $d = 2$ and $n=1$ in the absence of background contributions $U = W = 0$, in which case the solutions to the field equations~\eqref{EOM::pte}, subject to the boundary conditions~\eqref{asymp1} and~\eqref{asymp2}, are given by Bessel functions of the first kind,
\begin{align}
\left. \eta_{\pm} \right|_{U = W = 0} \propto J_0(k_{\pm} r).
\end{align}
This result is consistent with the decoupled form of the recurrence relations~\eqref{rec::1}. In light of this, it might be expected that the recurrence relations~\eqref{rec::1} give rise to a linear combination of Bessel functions, as we will see in the following.

\subsection{ Resummed solutions }
\label{sec:level3::3}

We have found that the series expansion of the perturbative scattering solutions leads to the recurrence relations~\eqref{rec::1}. In this subsection, we solve the recurrence relations to obtain compact analytical solutions. We first study the special case with $d=2$ and $n=1$, and then generalize the discussion to $d\ge2$ and $n \geq 2$. For convenience, in the case $n=1$, the superscripts are omitted.

$\bullet$ ~For the case $d=2$ and $n=1$, the recurrence relations (see Eq.~\eqref{rec::1}) can be rewritten in the following matrices:
\begin{align}
\begin{pmatrix}
c_{l+2} \\
d_{l+2}
\end{pmatrix} = \frac{1}{(l+2)^2} \begin{pmatrix}
U-k_+^2  & W \\
W  & U-k_-^2
\end{pmatrix}
\begin{pmatrix}
c_{l} \\
d_{l}
\end{pmatrix}.
\end{align}
For convenience, we introduce the two matrix variables:
\begin{align}
    \sigma_l = \frac{r^2}{(2l)^2} \gamma, \quad \gamma = 
\begin{pmatrix}
U-k_+^2  & W \\
W  & U-k_-^2
\end{pmatrix},
\end{align}
where $\gamma$ can be diagonalized as
\begin{align}
    \gamma = \lambda^{-1} \cdot \rho \cdot \lambda,
\end{align}
with $\rho=\diag (\rho_1,\rho_2)$ a diagonal matrix and $\lambda$ the corresponding matrix of eigenvectors. Using this notation, the solution can be compactly written as:
\begin{align}
\begin{pmatrix}
\eta_+(r) \\
\eta_-^*(r)
\end{pmatrix} = \chi_1
\begin{pmatrix}
c_{0} \\
d_{0}
\end{pmatrix},
\end{align}
where the matrix $\chi_j$ satisfies the recurrence relation:
\begin{align}
    \chi_j = \pmb{1} + \chi_{j+1} \cdot \sigma_{j}, \text{ for } j = 1,2,\cdots, 
\end{align}
and $\pmb{1}=\diag(1,1)$ is the $2\times2$ identity matrix. For finite $r$, we have
\begin{align}
    \lim_{l \to \infty} \sigma_l = \pmb{0}, \Rightarrow \lim_{l \to \infty} \chi_l = \pmb{1}.
\end{align}
Truncating the recursion at $l_{\max}$ by setting $\chi_{(l_{\max}+1)}=\pmb{1}$, the solution becomes 
\begin{align}
    \chi_1 = \lambda^{-1} \left( \pmb{1} + \left( \frac{r}{2} \right)^2 \rho + \cdots + \left( \prod_{l=1}^{l_{\max}} \frac{r}{2l} \right)^2 \rho^{l_{\max}} \right) \lambda .
\end{align}
In the limit $l_{\max} \to \infty$, the above series converges to the Bessel function expansion: 
\begin{align}
    \chi_1 = \lambda^{-1} 
\begin{pmatrix}
J_0 (\sqrt{-\rho_1} r)  & 0 \\
0  & J_0 (\sqrt{-\rho_2} r)
\end{pmatrix} \lambda.
\end{align}
Therefore, the solution and its radial derivative take the form 
\begin{align}
\label{imp3}
& ~~~~\begin{pmatrix}
\eta_+(r) &
\eta_-^*(r)
\end{pmatrix}^T \\ 
& = 
\lambda^{-1} \cdot {\rm diag}
\left(
J_0 (\sqrt{-\rho_1} r) , J_0 (\sqrt{-\rho_2} r) 
\right) \cdot \lambda \cdot
\begin{pmatrix}
c_{0} \\
d_{0}
\end{pmatrix}. \notag \\
&~~~~ - 
\begin{pmatrix}
\p_r \eta_+(r) & \p_r \eta_-^*(r)
\end{pmatrix}^T  \label{imp4}  \\
& \! = \lambda^{-1} {\rm diag} 
\left(
\sqrt{-\rho_1} J_1 (\sqrt{-\rho_1} r) , \sqrt{-\rho_2}  J_1 (\sqrt{-\rho_2} r) 
\right) \lambda
\begin{pmatrix}
c_{0} \\
d_{0}
\end{pmatrix}. \! \notag
\end{align}
Here, the matrix $\lambda$ is independent of the radial coordinate but depends on $\omega_Q$, $g$, $f_0$, and $\omega$. $\sqrt{-\rho_1}$ and $\sqrt{-\rho_2}$ can be interpreted as characteristic perturbative wavenumbers associated with the scattering into the $Q$-ball.

$\bullet$ ~For the case $d>2$ and $n=1$, introducing the transformation $\eta_\pm = \xi_\pm / r^{\delta} $, with $\delta=(d-2)/2$, the EoM can be rewritten as
\begin{align}
    \! \left( \p_r^2 + \frac{1}{r} \p_r \right) \xi_\pm + \left(k_\pm^2-U - \frac{\delta^2}{r^2} \right)\xi_\pm-W\xi_\mp^*=0. \!
\end{align}
From the boundary conditions in Eqs.~\eqref{asymp1} and \eqref{asymp2}, the corresponding boundary conditions for $\xi_\pm$ are given by
\begin{align}
    \lim_{r\to0} \frac{\xi_\pm}{r^\delta} & = \text{const.} , \\
    \lim_{r\to \infty} \xi_\pm & = \frac{k^{-\delta}_\pm}{\sqrt{k_\pm r}} \left( A_\pm e^{i k_\pm r} + B_\pm e^{-i k_\pm r} \right).
\end{align}
Then, by a procedure very similar to the $d=2$ and $n=1$ case, we find that the compact analytical solution for the case of  $d>2$ and $n=1$ is 
\begin{align}
\begin{pmatrix}
\xi_+ \\
\xi_-^*
\end{pmatrix} & = 
\lambda^{-1} 
\begin{pmatrix}
\frac{J_\delta (\sqrt{-\rho_1} r)}{(-\rho_1)^{\delta/2}}  & 0 \\
0  & \frac{J_\delta (\sqrt{-\rho_2} r)}{(-\rho_2)^{\delta/2}}
\end{pmatrix} \lambda
\begin{pmatrix}
c_{\delta} \\
d_{\delta}
\end{pmatrix}.
\end{align}
where $J_\delta$ denotes the Bessel function of the first kind of order $\delta$. 

$\bullet$ ~Now, we further consider the case of $n \geq 2$, where the superscripts are no longer omitted. In the second and subsequent spatial regions, the treatment of the series expansion must be modified slightly. To ensure continuity at the interfaces, the perturbative scattering solution must satisfy the matching conditions,
\begin{align}
    \eta^{(j)}_{\pm} (r_j)  = \eta^{(j+1)}_{\pm} (r_{j}), ~~
    \left. \p_r \eta_{\pm}^{(j)} \right|_{r=r_{j}} \!\!\! = \left. \p_r \eta_{\pm}^{(j+1)} \right|_{r=r_{j}},
    \label{mc::1}
\end{align}
where $j=1,2,\cdots,n-1$.
These continuity conditions provide four independent constraints, requiring four free parameters in the series expansion within each of the $n \geq 2$ regions. This marks a key difference from the analysis in the first region. To fully determine the solution, two additional linearly independent functions must be introduced in order to satisfy the matching conditions. It is straightforward to see that the necessary independent solution is provided by the Bessel function of the second kind (Neumann function). Consequently, in region $n\geq 2$, the general solution takes the form
\begin{align}
\label{imp5}
& \begin{pmatrix}
\xi_+^{(j)} \\
(\xi_-^{(j)})^*
\end{pmatrix} = 
\lambda^{-1} 
\begin{pmatrix}
\frac{J_\delta (\sqrt{-\rho_1} r)}{(-\rho_1)^{\delta/2}}  & 0 \\
0  & \frac{J_\delta (\sqrt{-\rho_2} r)}{(-\rho_2)^{\delta/2}}
\end{pmatrix} \lambda 
\begin{pmatrix}
c_{\delta}^{(j)} \\
d_{\delta}^{(j)}
\end{pmatrix} \notag \\
& ~~~~~\quad + \lambda^{-1} 
\begin{pmatrix}
\frac{N_\delta (\sqrt{-\rho_1} r)}{(-\rho_1)^{\delta/2}}  & 0 \\
0  & \frac{N_\delta (\sqrt{-\rho_2} r)}{(-\rho_2)^{\delta/2}}
\end{pmatrix} \lambda
\begin{pmatrix}
p_{\delta}^{(j)} \\
q_{\delta}^{(j)}
\end{pmatrix},
\end{align}
where $N_\delta(r)$ denotes the Neumann function, and the four independent coefficients $(c_{\delta}^{(j)},d_{\delta}^{(j)},p_{\delta}^{(j)},q_{\delta}^{(j)})$ are fixed by the four matching conditions (Eqs.~\eqref{mc::1}). Here we note that the coefficients $\rho_1,$ and $\rho_2$, as well as the matrix $\lambda$, depend on $U^{(j)},W^{(j)},k_+^{(j)}$ and $,k_-^{(j)}$.

\subsection{ Amplification factor }
\label{sec:level4}

Having derived the $Q$-ball solutions and their perturbations, we now turn to the analysis of the amplification behavior of various physical quantities. In particular, we will focus on the energy and energy flux associated with the ingoing and outgoing wave modes. 

Conservation of particle number plays a crucial role in this context \cite{Saffin:2022tub}, serving as a fundamental constraint in comparing the dynamics of the wave components. Explicitly, the ansatz~\eqref{ans::phi} exhibits both a scaling symmetry and a global $U(1)$ symmetry for the perturbative scattering solutions. The field equations~\eqref{EOM::pte} remain invariant under the transformation, 
\begin{align}
    (\eta_+,\eta_-) \to \alpha( e^{i\beta} \eta_+ , e^{-i\beta} \eta_-), 
\label{equ::scal}
\end{align}
where $\alpha$ and $\beta$ are real constants. It is simplest to see the implications of the global $U(1)$ symmetry by reconstructing the corresponding Lagrangian from the field equations~\eqref{EOM::pte}, which takes the form, 
\begin{align}
    \label{lag::er}
\mathcal{L}(\eta_{\pm}) = & \sum_{s=\pm} \left( - \eta_{s}^{\dagger} (\nabla^2 + k_s^2) \eta_{s} + U (\eta_{s}^{\dagger} \eta_s) \right) \notag \\
&  + W \left( \eta_+^\dagger \eta_-^\dagger + h.c. \right),
\end{align}
with $h.c.$ denoting the Hermitian conjugate. The associated Noether charge is then given by
\begin{align}
M_{\eta} =  i r^{d-1} \left(  \eta_{+}^{\dagger} \overleftrightarrow{\p_r} \eta_+  -  \eta_{-}^{\dagger} \overleftrightarrow{\p_r} \eta_- \right), 
\label{conspart}
\end{align}
where $\eta_{+}^\dagger \overleftrightarrow{\p_r} \eta_+ = \eta_+^\dagger \p_r \eta_+ - \p_r \eta_{+}^\dagger \eta_+$. That is, this quantity satisfies $\p_r M_{\eta} =0$, and thus $M_{\eta}$ is independent of $r$. From the boundary condition $M_\eta(r=0)=0$ (see Eq.~\eqref{asymp1}), we then conclude that $M_{\eta}\equiv0$ throughout the domain. 

The condition $M_{\eta}=0$ reflects particle number conservation in the scattering. By substituting the asymptotic forms $\eta_\pm^\infty$ into $M_\eta$ and integrating over a $(d-1)$-dimensional spherical shell region, the conservation law manifests as a balance between the ingoing and outgoing modes:
\begin{align}
\! N_c & \equiv \frac{|A_-|^2}{k_-^{d-2}} + \frac{|B_+|^2}{k_+^{d-2}}  = \frac{|B_-|^2}{k_-^{d-2}} + \frac{ |A_+|^2}{k_+^{d-2}} \\
 & = N^{in}_+ + N^{in}_- = N^{out}_+ + N^{out}_-, \!
 \label{cons2}
\end{align}
where $N_c$ denotes the total conserved particle number and we have, for clarity, defined the following particle numbers for the case $\omega>0$:
\begin{align}
    N^{in}_+ & = \frac{|B_+|^2}{k_+^{d-2}}, \; 
    N^{in}_- = \frac{|A_-|^2}{k_-^{d-2}}, \; \\
    N^{out}_+ & = \frac{|A_+|^2}{k_+^{d-2}}, \;
    N^{out}_- = \frac{|B_-|^2}{k_-^{d-2}}, \; 
\end{align}
where $A_-$ and $B_+$ represent the ingoing modes, and $A_+$ and $B_-$ the outgoing modes.
The conservation of particle number imposes the following constraint on the modes:
\begin{align}
\label{cons3}
    {N^{in}_+,N^{in}_-,N^{out}_+,N^{out}_-} \in [0,N_c].
\end{align}
Thus, the scattering process can be understood as a redistribution of particle numbers between modes, schematically represented as:
\begin{align}
(N^{in}_+ , N^{in}_- ) \xrightarrow{\text{scattering}}(N^{out}_+ , N^{out}_- ). 
\end{align}
It is important to emphasize that particle number conservation imposes constraints on the amplitudes of the ingoing and outgoing modes. Once a specific ingoing (or outgoing) configuration is chosen, thereby fixing the corresponding ingoing (or outgoing) particle number, the total particle number of the outgoing (or ingoing) configuration is fixed. The only freedom that remains lies in how this fixed particle number is distributed among the different outgoing (or ingoing) modes. Such redistribution may lead to superradiant behavior in other physical quantities. To further explore the implications of this effect, we turn to the energy-momentum tensor, which encompasses a range of physical quantities, including energy, energy flux, momentum, angular momentum, and stress. Among these, energy and energy flux are of particular interest. In what follows, we focus on these two quantities as representative examples to analyze superradiant behavior in the system.

Due to the exponential decay of the background field outside the core of the $Q$-ball, the dominant contributions to the energy and energy flux in this region arise from the perturbative scattering solutions. These contributions can be expressed explicitly as:
\begin{align}
    E & = T_{tt} = \left| \p_t \phi \right|^2 +  \left| \nabla \phi \right|^2 + |\phi|^2 + O(r^{-d}),\\
    P & = T_{rt} = \p_r \phi^* \p_t \phi + \p_t \phi^* \p_r \phi.
\end{align}
In the limit $r \to \infty$, the nonlinear terms from the potential decay rapidly and contribute only subleading corrections, which are neglected here. 

Substituting the asymptotic form of the field (see Eq.~\eqref{xi1}) into the above expressions and integrating over a $(d-1)$-spherical shell in the region from $r_a$ to $r_b$ as $r_a, r_b \to \infty$, the averaged energy and energy flux associated with the ingoing and outgoing modes are given by:
\begin{align}
    & E_\circledcirc  = \frac{1}{r_b-r_a} \int_{r_a}^{r_b} {\rm d}r r^{d-1} \left \langle {T_{tt}}  \right \rangle_{T\Omega}, \\
    & \! =  \frac{\omega_{+}^2}{k_+^{d-1}} \left( |A_+|^2 + |B_+|^2 \right) +  \frac{\omega_{-}^2}{k_-^{d-1}} \left( |A_-|^2 + |B_-|^2 \right), \! \notag \\
    & P_{rt}  = \frac{-1}{r_b-r_a} \int_{r_a}^{r_b} {\rm d}r r^{d-1} \left \langle {T_{rt}}  \right \rangle_{T\Omega}, \\
    & \! = \frac{\omega_+}{k_+^{d-2}} \left( - |A_+|^2 + |B_+|^2 \right) +  \frac{\omega_-}{k_-^{d-2}} \left( - |A_-|^2 + |B_-|^2 \right). \! \notag
\end{align}
where $\langle \cdot \rangle_{T\Omega}$ denotes the average over several temporal oscillations and over the entire $(d-1)$-sphere. Here, the shell region from $r_a$ to $r_b$ includes at least one full spatial oscillation of the longest wavelength. It is evident that both the ingoing and outgoing modes decompose into two distinct branches, characterized by the frequencies $\omega_+$ and $\omega_-$. Accordingly, each branch contributes separately to the energy and energy flux, given by:
\begin{align}
    E_+ = \frac{\omega_+^2}{k_+}, ~ E_- = \frac{\omega_-^2}{k_-} , \quad
    P_+ = \omega_+ , ~ P_- = -\omega_-. 
\end{align}
Note that All the four quantities above are positive.
Based on this identification, the amplification factors for energy and energy flux can be defined as follows:
\begin{align}
    \! \mathcal{A}_{tt} & =  \frac{ \frac{\omega_{+}^2}{k_+^{d-1}}  |A_+|^2 + \frac{\omega_{-}^2}{k_-^{d-1}} |B_-|^2 }{ \frac{\omega_{-}^2}{k_-^{d-1}}  |A_-|^2 +  \frac{\omega_{+}^2}{k_+^{d-1}} |B_+|^2 } 
    \\
    &= \frac{ E_+ N_+^{out} + E_- N_-^{out} }{ E_+ N_+^{in} + E_- N_-^{in} }  , \! \label{Ampc2} \\ 
    \!\mathcal{A}_{rt} & =  \frac{ \frac{\omega_+}{k_+^{d-2}}  |A_+|^2 + \frac{-\omega_-}{k_-^{d-2}} |B_-|^2 }{  \frac{-\omega_-}{k_-^{d-2}}  |A_-|^2 +  \frac{\omega_+}{k_+^{d-2}} |B_+|^2 } 
    \\
    &= \frac{ P_+ N_+^{out} + P_- N_-^{out} }{ P_+ N_+^{in} + P_- N_-^{in} }  . \!
    \label{Ampc3}
\end{align}
Each term in the numerators and denominators above is positive, as ensured by the condition $\omega>1+\omega_Q$, corresponding to propagating solutions.

We are particularly interested in two types of single ingoing mode configurations. Owing to the conservation of particle number and the scaling symmetry of the perturbative scattering solutions, we may, without loss of generality, normalize the total particle number to $N_c=1$. Under this normalization, the two single ingoing mode cases correspond to $N_-^{in}=0,~N_+^{in}=1$ and $N_-^{in}=1,~N_+^{in}=0$, respectively. For these two cases, the amplification factors for energy and energy flux take the following forms:
\begin{itemize}
    \item Case a: $N_-^{in}=0,~N_+^{in}=1$: 
    \begin{align}
    \label{atta}
    \mathcal{A}_{tt}^a & = \frac{E_-}{E_+} + \left( 1 - \frac{E_-}{E_+} \right) N_+^{out}, \\ 
    \mathcal{A}_{rt}^a & = \frac{P_-}{P_+} + \left( 1 - \frac{P_-}{P_+} \right) N_+^{out}; 
    \label{arta}
\end{align}
    \item Case b: $N_-^{in}=1,~N_+^{in}=0$: 
    \begin{align}
    \mathcal{A}_{tt}^b & = \frac{E_+}{E_-} + \left( 1 - \frac{E_+}{E_-} \right) N_-^{out}, \label{attb} \\ 
    \mathcal{A}_{rt}^b & = \frac{P_+}{P_-} + \left( 1 - \frac{P_+}{P_-} \right) N_-^{out}. 
    \label{artb}
\end{align}
\end{itemize}
Here, the superscripts $a$ and $b$ are used to distinguish the two distinct single ingoing mode configurations. The reason for retaining $N_+^{out}$ in the first case and $N_-^{out}$ in the second simplifies the subsequent analysis, as will become clear in the next section. Fundamentally, the amplification factors are governed by the redistribution of particle numbers between the two frequency branches during the scattering process. Given the normalization of the total ingoing particle number, the outgoing particle number can be interpreted as the retention/reflection rate of the single ingoing mode configuration, while the remaining component corresponds to the conversion/conversion rate into the other frequency branch. Accordingly, in the following analysis, we focus on computing the outgoing particle number $N_+^{out}$ in Case a and $N_-^{out}$ in Case b.

\subsection{Naive bounds on amplification factors}
\label{sec:naivebounds}

As mentioned, due to the particle number conservation, amplification of a physical quantity can occur if there exists a discrepancy between the corresponding quantities carried by the two modes. From Eq.~\eqref{Ampc2} (and similarly Eq.~\eqref{Ampc3}), we see that the amplification of the corresponding physical quantity can be interpreted as a weighted ratio of particle numbers, which, combining with Eq.~\eqref{cons2} and Eq.~\eqref{cons3}, provides some absolute constraints on the extent of amplification, which are given by:
\begin{align}
    \max (\! \mathcal{A}_{tt}) & = \frac{1}{\min (\! \mathcal{A}_{tt})}= \frac{\max(E_+,E_-)}{\min(E_+,E_-)}, \label{lim1} \\
    \max (\! \mathcal{A}_{rt}) & = \frac{1}{\min (\! \mathcal{A}_{rt})} = \frac{\max(P_+,P_-)}{\min(P_+,P_-)} = \frac{\omega_+}{-\omega_-}. \label{lim2}
\end{align}
If the ingoing or outgoing mode configuration is specified, the allowed range of the amplification factors can be further restricted to a narrower interval \cite{Zhang:2024ufh,Zhang:2025oud}. Note that in these estimates no information of the model or the background Q-ball is used, so these maximum values may not be reached in physical scattering in specific models. As we shall see later, upon obtaining analytical scattering solutions, much tighter bounds can be imposed (see Fig.~\ref{range1}).

\section{ Superradiance }
\label{sec:level5}

In this section, we extract the amplitudes of ingoing and outgoing modes by matching the asymptotic forms with the analytical solutions, from which the corresponding particle numbers and amplification factors can be obtained. We will derive explicit formulas for the amplification factors and clarify the role of different parameters in determining the amplification factors, starting with the ideal thin-wall limit in the case $n=1$ and $d=2$, and then generalizing the discussion to arbitrary dimension $d$ and general $n$.

\subsection{ Thin-wall limit }
\label{sec:level5::1}

In this subsection, we focus on the case $d=2$ and $n=1$, where the background field is divided into two regions: an inner region with a constant field value $f_Q=f_0$, and an outer region with $f_Q=0$. This profile corresponds to the ideal thin-wall limit. The interface between the two is located at $r_*$. To ensure the continuity and smoothness of the perturbative scattering solutions across the boundary, the following matching conditions between the analytical solutions and the asymptotic waves are imposed at $r=r_*$:
\begin{align}
    \eta_{\pm}(r_*) = \eta_{\pm}^\infty(r_*), \quad \left. \p_r\eta_{\pm}\right|_{r=r_*} = \left. \p_r \eta_{\pm}^\infty\right|_{r=r_*}.
\end{align}
By substituting Eqs.~\eqref{xi1} and \eqref{imp3} into the matching condition given above, we obtain the following relations:
\begin{align}
    u_1 ~ c_0 + u_2 ~ d_0 & \xlongequal{r=r_*}  v_1 ~ A_+ + v_2 ~ B_+, \notag \\
    u_3 ~ c_0 + u_4 ~ d_0 & \xlongequal{r=r_*} v_3 ~ B_-^* + v_4 ~ A_-^* , \notag \\
    \p u_1 ~ c_0 + \p u_2 ~ d_0  & \xlongequal{r=r_*} \p v_1 ~ A_+ + \p v_2 ~ B_+ , \notag \\
    \p u_3 ~ c_0 + \p u_4 ~ d_0  & \xlongequal{r=r_*} \p v_3 ~ B_-^* + \p v_4 ~ A_-^* ,
    \label{match1}
\end{align}
where $u_j,$ and $v_j$ for $j=1,2,3,4$ are the coefficient functions for the parameters $c_0,d_0,A_\pm,$ and $B_\pm$. The derivatives of the coefficients with respect to the radial coordinate are defined as
\begin{align}
    \p u_j  \equiv \frac{\p}{\p r} u_j(\omega_Q,\omega,r) , \quad
    \p v_j  \equiv \frac{\p}{\p r} v_j(\omega_Q,\omega,r) ,
\end{align}
where $j=1,2,3,4$. The symbol $\xlongequal{r=r_*}$ indicates that the two sides of the equations are equal only at $r=r_*$. These coefficients depend solely on the variables $\omega_Q,\omega$, and $r$. 
Since we are considering two regions, the superscripts originally distinguishing $c_0$ and $d_0$ between regions are omitted for clarity.  It is obvious that the system involves six variables: $c_0,d_0,A_\pm$, and $B_\pm$, and thus requires two additional constraints to yield a unique solution. These constraints can be imposed in various ways, such as fixing $c_0$ and $d_0$, or prescribing the amplitude of specific ingoing or outgoing modes. In this work, we choose to specify the ingoing modes. 

A natural choice is to consider a single ingoing mode, as discussed in the preceding section (see Eqs.~\eqref{atta}-\eqref{artb}). Imposing the corresponding constraints for the two distinct single ingoing mode configurations, the resulting solutions take the following forms:
\begin{itemize}
    \item Case a: $A_-=0, B_+=1$ : 
\begin{align}
    c_0 & = \frac{(v_1 \p v_2 - \p v_1 v_2) (v_3 \p u_4 - \p v_3 u_4)}{\pmb{v}_{13} \cdot \pmb{w} } , \\
    d_0 & = - \frac{(v_1 \p v_2 - \p v_1 v_2) (v_3 \p u_3 - \p v_3 u_3)}{\pmb{v}_{13} \cdot \pmb{w}} , \\
    A_+ & = - \frac{\pmb{v}_{23} \cdot \pmb{w}}{\pmb{v}_{13} \cdot \pmb{w}} , \label{eq:caap} \\
    B_-^* & = \frac{(v_1 \p v_2 - \p v_1 v_2) (u_3 \p u_4 - \p u_3 u_4)}{\pmb{v}_{13} \cdot \pmb{w}} , \label{eq:cabm}
\end{align}
    \item Case b: $A_-=1, B_+=0$ : 
\begin{align}
    c_0 & = - \frac{(v_1 \p u_2 - \p v_1 u_2) (v_3 \p v_4 - \p v_3 v_4)}{\pmb{v}_{13} \cdot \pmb{w} } , \\
    d_0 & = \frac{(v_1 \p u_1 - \p v_1 u_1) (v_3 \p v_4 - \p v_3 v_4)}{\pmb{v}_{13} \cdot \pmb{w}} , \\
    A_+ & = - \frac{(u_1 \p u_2 - \p u_1 u_2) (v_3 \p v_4 - \p v_3 v_4)}{\pmb{v}_{13} \cdot \pmb{w}} , \\
    B_-^* & = - \frac{\pmb{v}_{14} \cdot \pmb{w}}{\pmb{v}_{13} \cdot \pmb{w}} .
\end{align}
\end{itemize}
Here, we have defined auxiliary vectors for simplifying the expressions:
\begin{align}
    \pmb{w} & = (w_1,w_2,w_3,w_4), \\
    w_1 & = \p u_1 \p u_4 - \p u_2 \p u_3, \quad
    w_2 = \p u_1  u_4 - \p u_2  u_3, \notag \\
    w_3 & =  u_1 \p u_4 -  u_2 \p u_3, \qquad
    w_4 =  u_1  u_4 -  u_2  u_3, \notag \\
    \pmb{v}_{jk} & = (v_j v_k, - v_j \p v_k, - \p v_j v_k , \p v_j \p v_k), \\
    & \qquad \text{ for } j,k=1,2,3,4. \notag
\end{align}
Due to the specific structure of the asymptotic solutions, we have the useful relations:
\begin{align}
    v_1^* &= v_2 , \quad \frac{\p v_1}{v_1} = \left( \frac{\p v_2}{v_2} \right)^* = \left( i k_+ - \frac{d-1}{2r} \right), \notag \\ 
    v_3^* &= v_4,  \quad \frac{\p v_3}{v_3} = \left( \frac{\p v_4}{v_4} \right)^* = \left( i k_- - \frac{d-1}{2r} \right),
\end{align}
which leads to the following symmetry constraints among the auxiliary vectors:
\begin{align}
    \pmb{v}_{13} = \pmb{v}_{14}^* = \pmb{v}_{23}^* = \pmb{v}_{24}.
    \label{vequ}
\end{align}
Notice that while $\boldsymbol{v}_{ij}$ necessarily contains an imaginary part, all components of the vector $\pmb{w}$ are real-valued in the case of propagating scattering waves.

For two complex vectors $\pmb{a},\pmb{b}$ of the same dimension, the inner product can be expressed as:
\begin{align}
\boldsymbol{a} \cdot \boldsymbol{b}^* = |\boldsymbol{a}|  |\boldsymbol{b}|  \cos \theta_H(\pmb{a},\pmb{b}) ~ \exp (i \theta_K),
\end{align}
where $\theta_H(\pmb{a},\pmb{b})$ is the Hermitian angle between the vectors ($\pmb{a}$ and $\pmb{b}$), and $\theta_K$ is the pseudo-angle. In our analysis, only the Hermitian angle contributes to the physical quantity of interest. Therefore, the outgoing particle numbers $N_+^{out}$ for Case a and $N_-^{out}$ for Case b take the following forms:
\begin{align}
    N_+^{out} = |A_+|^2 = \frac{\cos^2 \theta_H(\pmb{v}_{23},\pmb{w})}{\cos^2 \theta_H(\pmb{v}_{13},\pmb{w})} = N_-^{out}, \label{nout} 
\end{align}
where we have used Eq.~\eqref{vequ}. From this relation, we can further obtain the following nontrivial constraints:
\begin{align}
    \left( 1 - N_+^{out} \right)^2 & = \left( \mathcal{A}_{tt}^{a} - N_+^{out} \right) \left( \mathcal{A}_{tt}^{b} - N_+^{out} \right), \notag \\
    & = \left( \mathcal{A}_{rt}^{a} - N_+^{out} \right) \left( \mathcal{A}_{rt}^{b} - N_+^{out} \right). 
    \label{imp1}
\end{align}
These constraints explicitly link the amplification factors of the two distinct single ingoing mode configurations. Moreover, according to Eq.~\eqref{cons2}, together with the normalization of the ingoing modes, the outgoing particle number is constrained to the range $[0,1]$. (If the total particle number $N_c$ is not normalized to be 1, $N^{out}_+$ in Eq.~(\ref{imp1}) should be replaced by  $N^{out}_+/N_c$.) We note that the explicit form of Eq.~\eqref{nout} in the ideal thin-wall limit with $d=2$ is determined by five parameters: $\omega_Q$, $g$, $f_0$, $r_*$, and $\omega$. 

As mentioned, the amplification factors depend only on the Hermitian angles between certain auxiliary vectors, and not on their magnitudes. This observation allows us to simplify the auxiliary vectors by normalizing them as follows:
\begin{align}
\label{approxv}
    \overline{\pmb{v}}_{jk} = \frac{\pmb{v}_{jk}}{v_j v_k} = \left(1, -\frac{\p v_k}{v_k} , -\frac{\p v_j}{v_j} , \frac{\p v_j \p v_k}{v_j v_k} \right).
\end{align}

Based on the above perturbative scattering solutions Eq.~\eqref{imp3}, the coefficients associated with $c_0$ and $d_0$ are found to take the following form:
\begin{align}
    u_1 & = \frac{1}{2} J_0^{+} + \frac{\omega_Q \omega}{\sqrt{W^2 + 4\omega_Q^2 \omega^2}} J_0^{-}, \notag \\
    u_2 & = u_3 =  - \frac{W}{2 \sqrt{W^2 + 4\omega_Q^2 \omega^2}} J_0^{-}, \notag \\
    u_4 & = \frac{1}{2} J_0^{+} - \frac{\omega_Q \omega}{\sqrt{W^2 + 4\omega_Q^2 \omega^2}} J_0^{-},
\end{align}
where
\begin{align}
    J_0^{\pm} & = J_0(\sqrt{-\rho_1} r) \pm J_0(\sqrt{-\rho_2} r), \\
    -\rho_1 & = \omega_Q^2 + \omega^2 + \sqrt{W^2 + 4\omega_Q^2 \omega^2} - (1 + U), \\
    -\rho_2 & = \omega_Q^2 + \omega^2 -  \sqrt{W^2 + 4\omega_Q^2 \omega^2} - (1 + U).
\end{align}
Finally, the components of the auxiliary vector $\pmb{w}$ are given by
\bal
    w_1 & = \sqrt{\rho_1 \rho_2} J_1 (\sqrt{-\rho_1} r) J_1 (\sqrt{-\rho_2} r), \label{uc1}
    \\
    w_2 & = \frac{1}{2} \left( J_0(\sqrt{-\rho_1} r) J_0(\sqrt{-\rho_2} r) \right)^\prime , \label{uc2}
    \\
    &~~~+ \frac{\omega_Q \omega}{\sqrt{W^2 + 4\omega_Q^2 \omega^2}} J_0^2 (\sqrt{-\rho_2} r) \left( \frac{J_0 (\sqrt{-\rho_1}r}{J_0 (\sqrt{-\rho_2}r} \right)^\prime  , \nonumber  \\
    w_3 & = \frac{1}{2} \left( J_0(\sqrt{-\rho_1} r) J_0(\sqrt{-\rho_2} r) \right)^\prime \label{uc3} 
    \\
    &~~~ - \frac{\omega_Q \omega}{\sqrt{W^2 + 4\omega_Q^2 \omega^2}} J_0^2 (\sqrt{-\rho_2} r) \left( \frac{J_0 (\sqrt{-\rho_1}r}{J_0 (\sqrt{-\rho_2}r} \right)^\prime, \nonumber \\
    w_4 &= J_0 (\sqrt{-\rho_1} r) J_0 (\sqrt{-\rho_2} r) .
\eal
With the explicit form of the auxiliary vectors, by substituting them into Eq.~\eqref{nout}, the outgoing particle number—and consequently the amplification factors Eqs.~\eqref{atta}-\eqref{artb}—can be determined.

\begin{figure*}
	\centering
        \includegraphics[height=6.24cm]{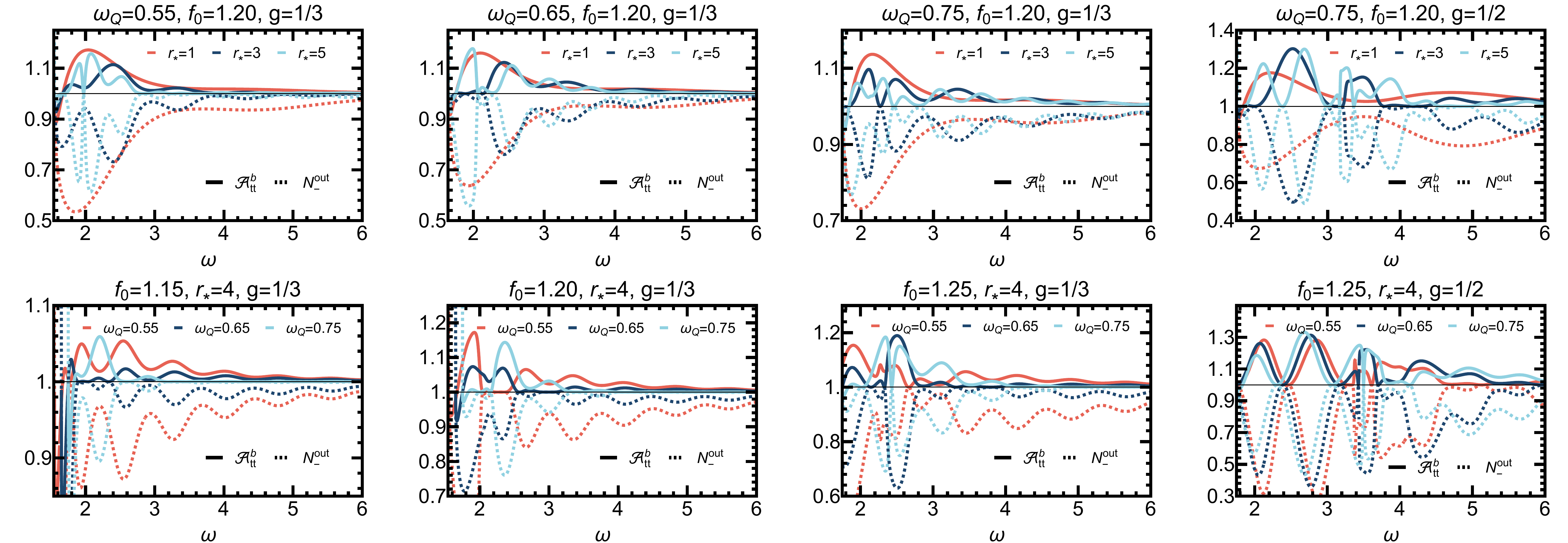}
	\caption{
    Energy amplification factor $\mathcal{A}_{tt}^b$ and outgoing particle number $N^{out}_-$ with different parameters $\omega_Q , f_0 , r_*$ and $g$, from the full analytical results. The extrema of the outgoing particle number and the amplification factors almost coincide, and the value of the thin-wall location $r_*$ turns to be the key parameter that controls the number of peaks in the amplification factors.
    }
 \label{exac}
\end{figure*}

In Fig.~\ref{exac}, we display the energy amplification factor and the exact outgoing particle number (Eq.~\eqref{nout}) for various parameter choices of $\omega_Q$, $f_0$, $r_*$, and $g$. It is found that the thin-wall location $r_*$ plays the central role in controlling the number of peaks in the amplification factors. The interesting feature of the figure is that the extrema of the outgoing particle number and the amplification factors almost coincide. However, they do not overlap exactly, as detailed in Appendix~\ref{sec:levela}. 

In the next subsection, by taking the large-$r_*$ limit, we will show that the outgoing particle number takes the form of a rational function of sinusoidal functions. Consequently, as $\omega$ varies, the outgoing particle number oscillates, especially for large $r_*$. The correlation between the outgoing particle number and the amplification factors can then be understood intuitively from Eqs.~\eqref{atta}–\eqref{artb}, where the two quantities are linearly related. Compared with the slow variations of $E_-/E_+$ or $P_-/P_+$, the oscillations of the outgoing particle number dominate the variations of the amplification factors near their extrema.

\begin{figure*}
    \centering
    \includegraphics[height=4.0cm]{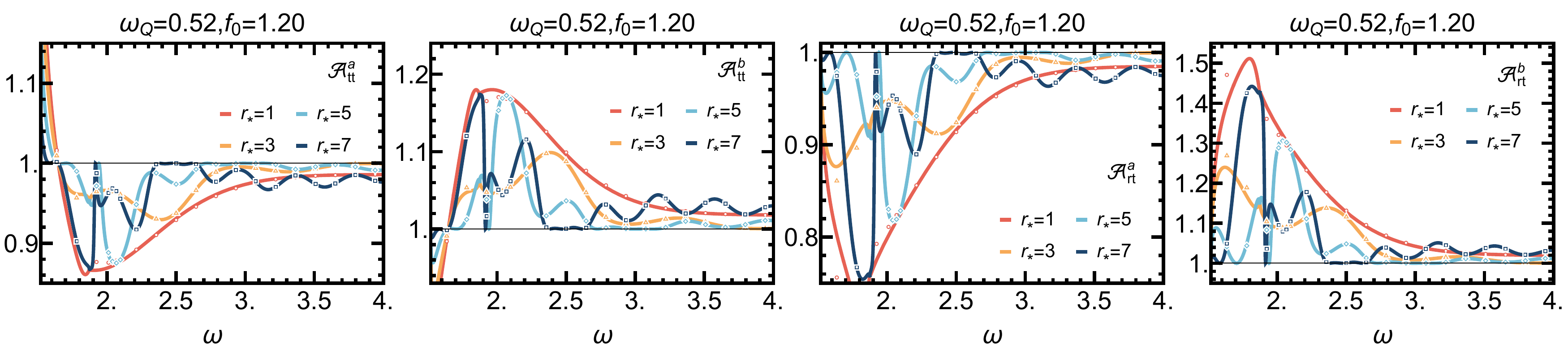}
    \includegraphics[height=8.6cm]{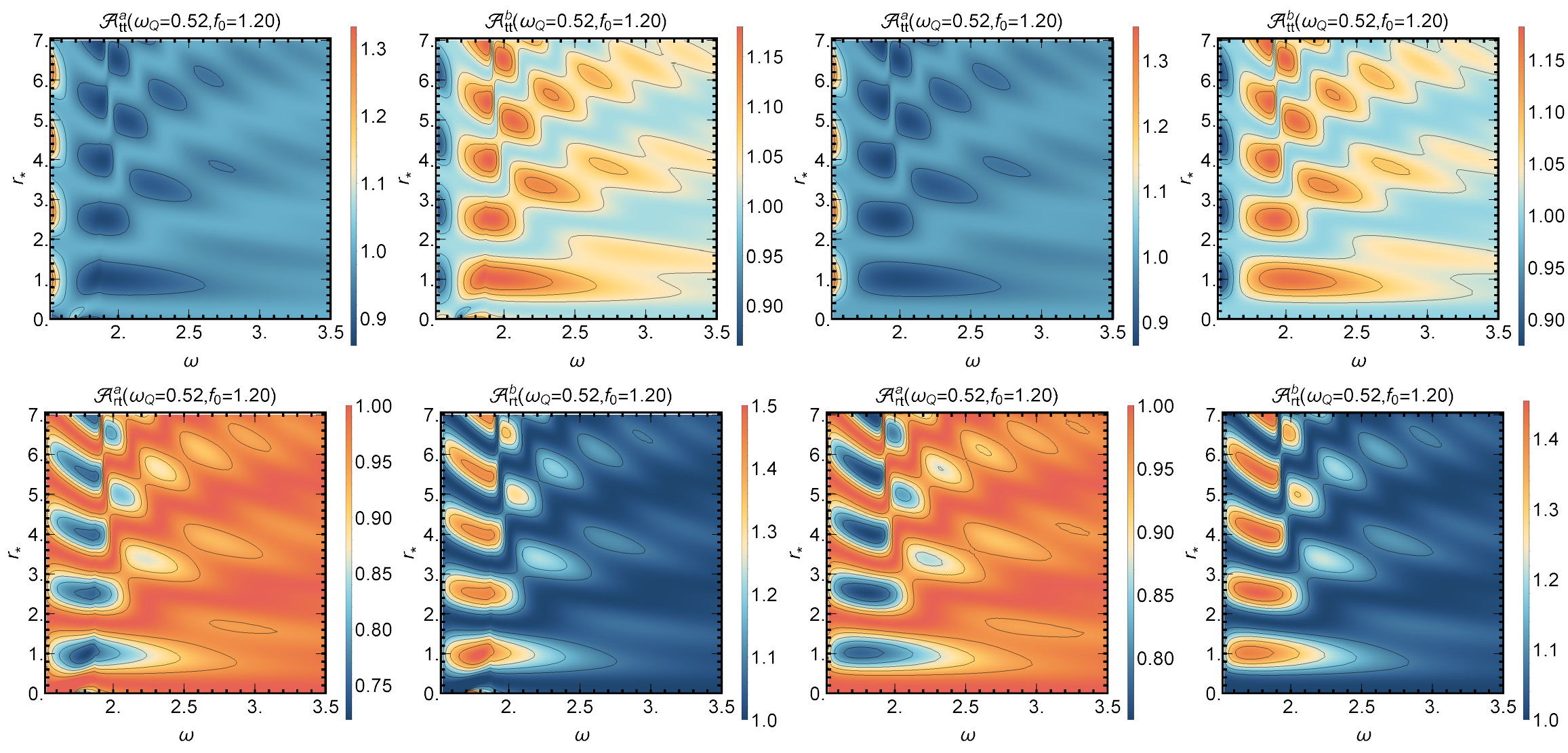}
    \caption{
    Amplification factors of energy $\mathcal{A}_{tt}^a,\mathcal{A}_{tt}^b$ and energy flux $\mathcal{A}_{rt}^a,\mathcal{A}_{rt}^b$ obtained from Eqs.~\eqref{atta}-\eqref{artb} with $\omega_Q=0.52$ and $g=1/3$ for a thin-wall Q-ball ($n=1$). 
    The top row shows solid lines for $N^{out}_\pm$ obtained from Eq.~\eqref{app1}, while the dotted lines indicate the numerical results, overlapping with the case where $N^{out}_\pm$ is obtained from Eq.~\eqref{nout}. In the bottom two rows, the left two columns present the approximate $N^{out}_\pm$ from Eq.~\eqref{app1}, which is valid for large $\omega$ or large $r_*$, whereas the right two columns display the $N^{out}_\pm$ obtained from Eq.~\eqref{nout}, which remain valid for all $\omega$ and $r_*$.
    }
 \label{comp}
\end{figure*}

\subsection{ Large $Q$-ball }
\label{sec:level5::2}

Let us now look at the large $r_*$ limit of the thin-wall analytical results, which allows us to extract simpler and more intuitive forms for the amplification factors.

In the large $r$ limit, the argument of the Bessel function $\sqrt{-\rho_i} r$ becomes large. As is well known, the Bessel function admits the following asymptotic form as $z\to \infty$:
\begin{align}
    J_\delta (z) \sim \sqrt{\frac{2}{\pi z}} \cos \left( z - \frac{\delta}{2}\pi - \frac{\pi}{4} \right).
\end{align}
Therefore, we neglect the subleading $O(1/z)$ terms (including those in Eq.~\eqref{approxv}), and the solution can be well approximated by Eqs.~\eqref{app1}-\eqref{app3}. Substituting the obtained outgoing particle number (Eq.~\eqref{app1}) into Eqs.~\eqref{atta}–\eqref{artb}, the corresponding amplification factors can be determined. 

Note that since $\sqrt{-\rho_i}\sim \omega$, the same limit of the Bessel function can be effected by taking $\omega$ large, which explains that in the plots the approximation is rather accurate in the large $\omega$ region even for a small $r_*$. However, later, we will take the large $\omega$ limit on top of the large $r_*$ limit, which allows us to further simplify the analytical results.

\begin{widetext}
\begin{align}
    N_+^{out} & = \left. \frac{ \left[ C_+ \cos( \sigma_- r) - C_- \sin( \sigma_+ r)  \right]^2 + \left[ (-k_- D_+ + k_+ D_-) \cos (\sigma_+ r) + ( k_- F_+ - k_+ F_- ) \sin (\sigma_- r) \right]^2 }{ \left[ C_- \cos( \sigma_- r) - C_+ \sin( \sigma_+ r)  \right]^2 + \left[ (k_- D_+ + k_+ D_-) \cos (\sigma_+ r) - ( k_- F_+ + k_+ F_- ) \sin (\sigma_- r) \right]^2 } \right|_{r=r_*} \label{app1} , \\
    \sigma_\pm & = \sqrt{-\rho_1} \pm \sqrt{-\rho_2}, \quad C_\pm = \sqrt{\rho_1 \rho_2} \pm k_+ k_- , \label{app2} \\
    D_\pm & = \frac{1}{2} \left( \sigma_+ \pm \frac{ 1 }{\sqrt{1+ W^2/(4\omega_Q^2 \omega^2) }} \sigma_- \right) , \quad
    F_\pm = \frac{1}{2} \left( \sigma_- \pm \frac{ 1 }{\sqrt{1+ W^2/(4\omega_Q^2 \omega^2) }} \sigma_+ \right). \label{app3}
\end{align}
\end{widetext} 

The sinusoidal form of Eq.~\eqref{app1} implies that the outgoing particle number oscillates rapidly with $\si_\pm$ for large $r_*$. By Eqs.~\eqref{atta}-\eqref{artb}, the amplification factors follow suit. Notice that $\si_+\sim 2\omega$ and $\si_-\sim 2\omega_Q$. This explains the previous observation that the amplification factors oscillate more rapidly for a larger $Q$-ball, which will become clearer when we take the large $\oi$ limit later.

Next, let us rewrite the frequency as $\omega=1+\omega_Q+\epsilon$ and $0<\epsilon \ll 1$, and examine the limiting behavior of the outgoing particle number. We find that
\begin{align}
    \lim_{\epsilon \to 0} N_{\pm}^{out} = 1 = \lim_{\epsilon \to \infty} N_{\pm}^{out}.
\end{align}
Substituting this into the amplification factors associated with the two independent ingoing modes (see Eqs.~\eqref{atta}-\eqref{artb}), we obtain the following limits:
\begin{align}
    \lim_{\epsilon \to 0} \mathcal{A}_{tt}^a & = const>1, \\
    \lim_{\epsilon \to 0} \mathcal{A}_{tt}^b & = \lim_{\epsilon \to 0} \mathcal{A}_{rt}^a = \lim_{\epsilon \to 0} \mathcal{A}_{rt}^b =1.
\end{align}
This result shows that the system does not exhibit divergent amplification in the threshold limit. Moreover, as $\omega \to \infty $, all amplification factors approach $1$.

\begin{figure}
    \centering
    \includegraphics[height=7.2cm]{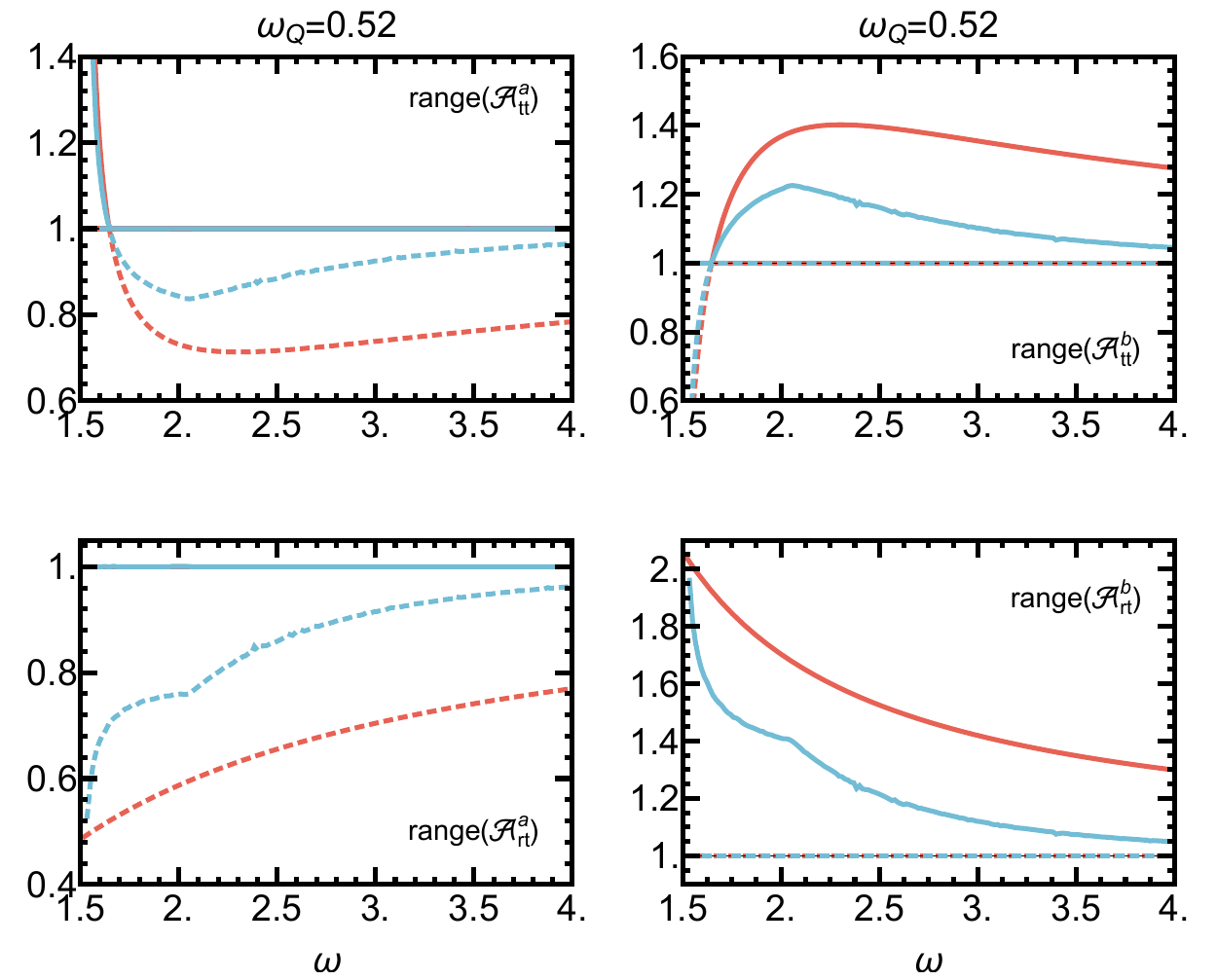}
    \caption{
    Limits of the energy and energy flux amplification factors for $\omega_Q=0.52$ and $g=1/3$, with $f_0\in[f_z,f_{\rm max})$ and $r_*\in[3,50]$, for the thin-wall $Q$-ball ($n=1$). The red lines indicate the limits given by Eqs.~\eqref{lim1} and \eqref{lim2}, while the blue lines correspond to those from Eqs.~\eqref{app1}. Solid lines represent the upper bounds, and dotted lines denote the lower bounds.
    }
 \label{range1}
\end{figure}

In Fig.~\ref{comp}, we compare the amplification factors computed using the approximate expression Eq.~\eqref{app1}, valid for large $\omega$ or $r_*$, with the exact results obtained either analytically from Eq.~\eqref{nout} or numerically from the perturbation equation. The two exact results are in perfect agreement. In the top row, different colors correspond to different matching points $r_*$. For $r_*\ge3$, the approximate and exact results exhibit excellent agreement, with the two sets of curves essentially overlapping. As expected, for $r_*=1$, the approximation is also very good for large $\omega$. In the bottom two rows of Fig.~\ref{comp}, we present the dependence of the approximate and exact expressions on the variables $\omega$ and $r_*$. The left two columns display the approximate results obtained from Eq.~\eqref{app1}, while the right two columns present the exact results obtained from Eq.~\eqref{nout}. 

Let us now refine the bounds on the amplification factors (Eqs.~\eqref{lim1} and \eqref{lim2}) with the large-$r_*$ solution. Note that for a large $Q$-ball, the amplification factors depend only on five parameters, $\omega_Q,g,f_0,r_*$, and $\omega$, with the allowed ranges of the first three parameters specified in Eq.~\eqref{pararange}. By varying these parameters, we can determine the bounds on the amplification factors. In Fig.~\ref{range1}, we consider the case $\omega_Q=0.52,g=1/3,f_0\in[f_z,f_{\max})$ and $r_*\in[3,50]$, where Eq.~\eqref{app1} is valid for large $r_*$. We see that the naive bounds obtained in Section \ref{sec:naivebounds} are indeed rather weak in nature. (Beyond the large-$r_*$ limit, the bounds on the amplification factors may be obtained by employing the exact outgoing particle number given in Eq.~\eqref{nout}.)

\subsection{Large $\oi$ limit of large $Q$-ball}

To obtain even simpler and more illuminating analytical results for the amplification factors, we can further take a large $\omega$ limit ($\omega \gg W/(2\omega_Q)$), on top of the large $r_*$ limit for the outgoing particle number $N_+^{out}$ ($N_+^{out}=N_-^{out}$ for single ingoing mode scenarios, cf.~\eqref{nout}). Note that, in the large $\oi$ limit, $E_+/E_-\sim P_+/P_-\sim 1$, so from Eq.~\eqref{attb} and \eqref{artb} we see that the variations of the amplification factors are dominated by those of $N_+^{out}$.

Let us now specify how to take a (partial) large $\omega$ limit. First of all, in the large $\omega$ limit, we recognize that $k_+$, $k_-$ and $\sqrt{-\rho_i}$ are of the order of $\omega$: $k_+\sim k_-\sim \sqrt{-\rho_i}\sim\omega$. Also, since $\cos(\si_\pm r)$ and $\sin(\si_\pm r)$ are bounded and we are interested in the oscillation behaviors of the amplification factors, we shall refrain from expanding them in terms of large $\oi$. Then, we note that an interesting fact about the denominator of Eq.~\eqref{app1} is that its leading large $\omega$ behavior goes like $C_+^2$, which does not contain the sinusoidal oscillation with $r_*$. Therefore, taking the leading $C_+^2$ term for the denominator in Eq.~\eqref{app1}, the large $\oi$ limit gives us
\begin{align}
\label{appn2}
    N^{out}_{+(2)} & = 1 + \kappa_1 \sin(\sigma_- r_*)^2 \notag \\
    & \quad\quad + \kappa_2 \sin(\sigma_- r_*) \cos( \sigma_+ r_*) \notag\\
    & \quad\quad + \kappa_3 \cos(\sigma_+ r_*)^2, 
    \\
    &\simeq 1  -\frac{W^2}{4\omega_Q^2\omega^2} \sin(2r_*\,  \omega_Q + \varphi_-)^2 \notag \\
    & \quad\quad + \frac{W^2}{2\omega_Q\omega^3} \sin(2 r_*\, \omega_Q + \varphi_- ) \cos( 2 r_* \cdot \omega + \varphi_+ ) \nonumber
    \\
    &\quad\quad +O(\oi^{-4})
    \label{centralRes}
\end{align}
where again $r_*$ is the radius of the thin-wall $Q$-ball and the coefficients $\kappa_1$ and $\kappa_2$ are defined as
\begin{align}
    \kappa_1 & =-\frac{4k_+k_-}{C_+^2}F_+F_- - 1 \simeq -\frac{W^2}{4\omega_Q^2\omega^2}, \notag\\
    \kappa_2 & =\frac{4k_+k_-}{C_+^2}(D_+F_-+D_-F_+) \simeq \frac{W^2}{2\omega_Q\omega^3}, \notag \\
    \kappa_3 & =-\frac{4k_+k_-}{C_+^2}D_+D_- + 1 \simeq \frac{U^2 - W^2}{4\omega^4},
\end{align}

\begin{figure}[H]
	\centering
	\includegraphics[height=7.2cm]{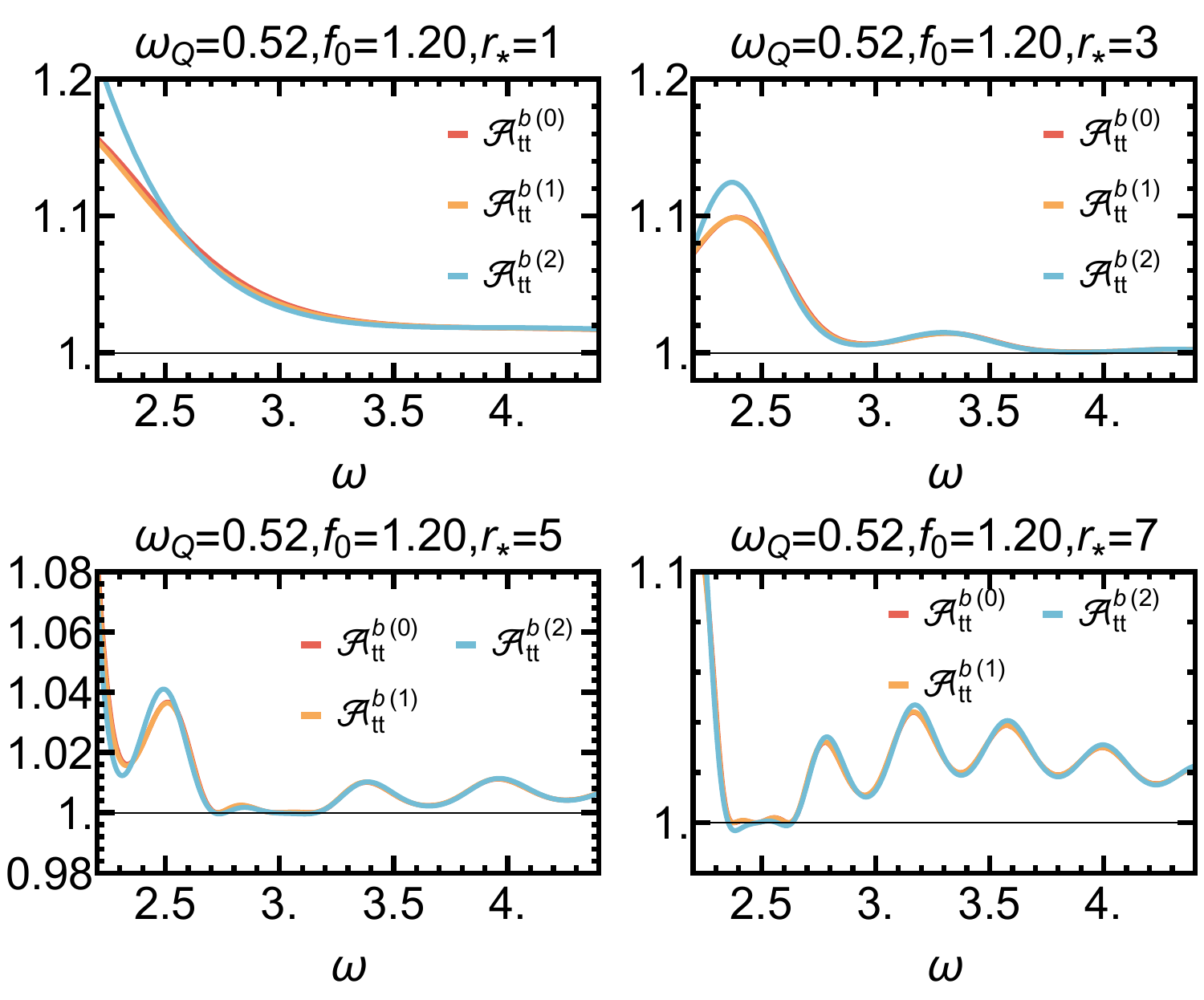}
	\caption{
     Comparisons of the energy amplification factor in different approximations. ${\cal A}^{b(0)}_{\rm tt}$ is the full analytical result, ${\cal A}^{b(1)}_{\rm tt}$ is the large $r_*$ limit, and ${\cal A}^{b(2)}_{\rm tt}$ is given by Eq.~\eqref{appn2}. 
    }
 \label{comp1}
\end{figure}

To get Eq.~\eqref{centralRes}, note that in the large $\omega$ limit, $\sigma_+$ and $\sigma_-$ become $2\omega$ and $2\omega_Q$ respectively to leading order. For a large $r_*$, it happens that the next-to-leading order contribute a sizable phase 
\begin{align}
\varphi_+ & = - \frac{1+U}{\omega} r_* , \notag \\
\varphi_- & = \frac{ W^2 + 4 ( 1 + U + 2\omega^2 ) \omega_Q^2 }{4\omega_Q \omega^2} r_* .
\end{align}
Plugging these into Eq.~\eqref{attb} and using $  E_+/E_-\simeq 1+2\omega_Q/\omega$, we find that
\begin{align}
\label{centralRes2}
    \mathcal{A}^b_{tt} & \simeq 1 + \frac{W^2}{2\omega_Q\omega^3} \sin (2 r_* \cdot \omega_Q + \varphi_- )^2  \\
    & -\frac{W^2}{\omega^4} \sin(2 r_* \cdot \omega_Q + \varphi_- ) \cos( 2 r_* \cdot \omega + \varphi_+ ) + O(\oi^{-5}). \notag 
\end{align}

\begin{figure}[H]
	\centering
	\includegraphics[height=5.5cm]{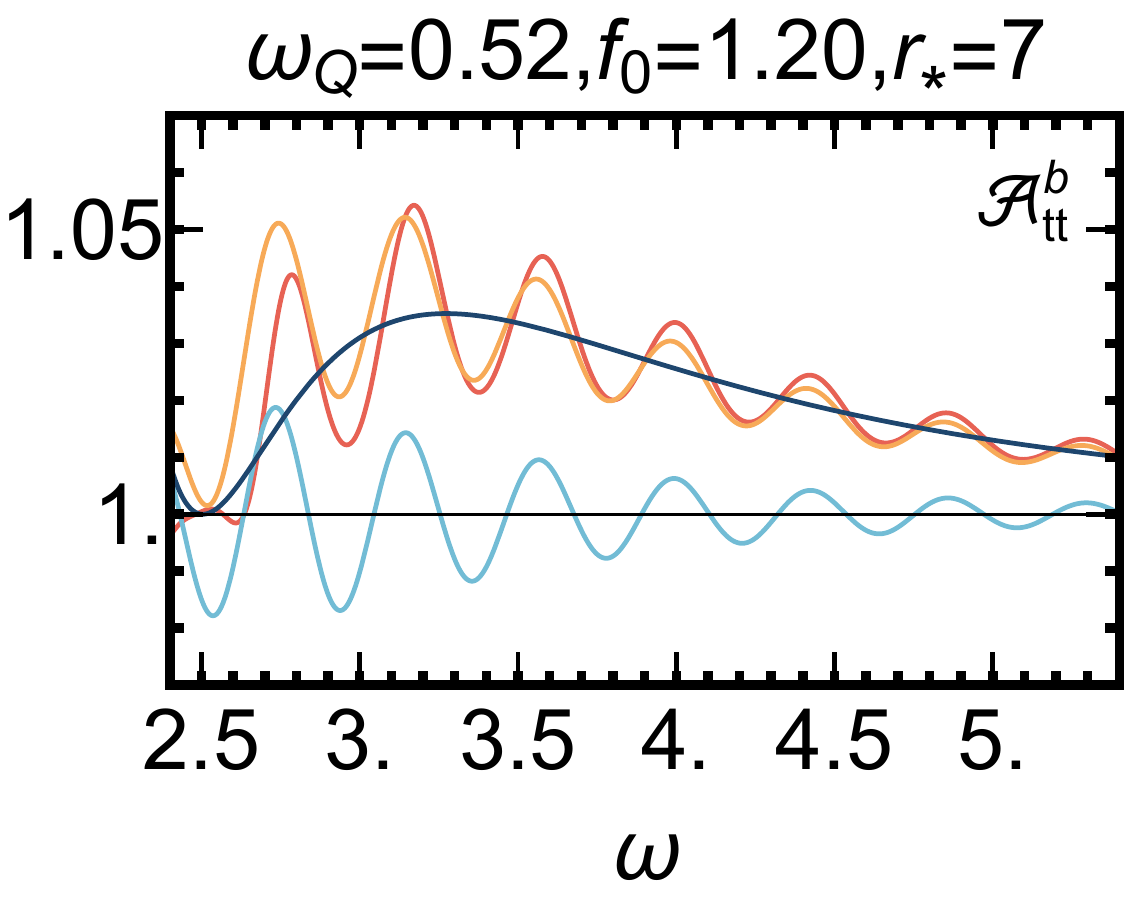}
	\caption{
     Main contributions to the energy amplification factor. The red curve represents the fairly accurate result of substituting Eq.~\eqref{appn2} into Eq.~\eqref{attb}. The orange curve, given by Eq.~\eqref{centralRes2}, consists of two main parts:  the dark blue curve depicts the base behavior of $\sin(2r_* \cdot \omega_Q + \varphi_-)^2/\oi^3$ plus 1, and the light blue curve represents the damped oscillations from $\sin(2 r_* \cdot \omega_Q + \varphi_- ) \cos( 2 r_* \cdot \omega + \varphi_+ )/\oi^4$ (up-shifted by 1 for easier visualization).  
    }
 \label{comp2}
\end{figure}

In Fig.~\ref{comp1}, we compare the amplification factors obtained from the thin-wall limit, the double limit of large $r_*$ and $\oi$, and the full analytical result. In the large $\omega$ regime, the approximation $N^{out}_{+(2)}$ demonstrates excellent agreement with the full result. This confirms that Eq.~\eqref{appn2} offers a relatively accurate and analytically tractable expression for the conversion rate. The form of Eq.~\eqref{centralRes} is particularly suggestive for explaining the origin of the oscillating, multi-peak structure in the spectra of the amplification factors, as shown in Fig.~\ref{comp2}: $1 + W^2 \sin (2 r_* \cdot \omega_Q + \varphi_- )^2 / (2\omega_Q\omega^3)$ provides the base behavior of the amplification factor, which is damped by $\oi^3$ and modulated by the $Q$-ball's frequency $\oi_Q$ and the $Q$-ball size $2r_*$; on top of that are damped oscillations $-W^2 \sin(2 r_* \cdot \omega_Q + + \varphi_-) \cos( 2 r_* \cdot \omega + + \varphi_+ )/\omega^4$, whose frequency is determined by the $Q$-ball size and which is again modulated by the $Q$-ball's frequency and the $Q$-ball size.

\subsection{ General case }
\label{sec:level5::4}

We have focused on the thin-wall $Q$-ball in the previous subsections. Now, we extend the analysis to the case of a general $Q$-ball that needs to be approximated with multiple piecewise steps ($n \geq 2$). 

For a clear presentation, we rewrite the perturbative scattering solution as
\begin{align}
\! \eta^{(j)}_{+}(r) & =
\begin{pmatrix}
 u^{(j)}_{c+} (r) & u^{(j)}_{d+} (r) & u^{(j)}_{p+} (r) & u^{(j)}_{q+} (r)
\end{pmatrix} 
\cdot \pmb{c}^{(j)} \! \notag , \\
\! \eta^{(j)}_{-}(r)^* & =
\begin{pmatrix}
 u^{(j)}_{c-} (r) & u^{(j)}_{d-} (r) & u^{(j)}_{p-} (r) & u^{(j)}_{q-} (r)
\end{pmatrix} 
\cdot \pmb{c}^{(j)}  \! \notag , \\
\pmb{c}^{(j)} & = 
\begin{pmatrix}
 c^{(j)}_{\delta} & d^{(j)}_{\delta} & p^{(j)}_{\delta} & q^{(j)}_{\delta}
\end{pmatrix}^{T},
\end{align}
where $j=1,2,\cdots,n+1$, and define their derivatives with respect to $r$ as
\begin{align}
    \p u^{(j)}_\#(r_\ell) = \left. \frac{\p}{\p r} u^{(j)}_\#(r) \right|_{r=r_\ell}, 
\end{align}
where $\# \in \{c_\pm,d_\pm,p_\pm,q_\pm\}$ and $r_\ell$ are the matching points. For the special cases $j=1$ and $j=n+1$, the coefficients take the form  
\begin{align}
    u^{(1)}_{p+} & = u^{(1)}_{q+} = u^{(1)}_{p-} = u^{(1)}_{q-} =0, \\
    u^{(n+1)}_{p+} & = u^{(n+1)}_{q+} = u^{(n+1)}_{c-} = u^{(n+1)}_{d-} = 0, \notag \\
    u^{(n+1)}_{c+} & = v_1(r), ~ u^{(n+1)}_{d+}  = v_2(r), \notag \\
    u^{(n+1)}_{p-} & = v_3(r), ~ u^{(n+1)}_{q-}  = v_4(r), \\
    c_{\delta}^{(n+1)} & = A_+, ~ d_{\delta}^{(n+1)} = B_+ , \notag \\
    p_{\delta}^{(n+1)} & = B_-^*, ~ q_{\delta}^{(n+1)} = A_-^*,
\end{align}
where 
\begin{align}
    v_1 (r) = v_2^* (r) = \frac{1}{(k_+ r)^{(d-1)/2}} e^{ik_+ r} , \notag \\
    v_3 (r) = v_4^* (r) = \frac{1}{(k_- r)^{(d-1)/2}} e^{ik_- r} .
\end{align}
In the general case, the matching conditions can be formulated as the linear equations, 
\begin{widetext}
    \begin{align}
    \left. 
        \begin{pmatrix}
 u^{(j)}_{c+} & u^{(j)}_{d+} & u^{(j)}_{p+} & u^{(j)}_{q+} & -u^{(j+1)}_{c+} & -u^{(j+1)}_{d+} & -u^{(j+1)}_{p+} & -u^{(j+1)}_{q+} \\
 u^{(j)}_{c-} & u^{(j)}_{d-} & u^{(j)}_{p-} & u^{(j)}_{q-} & -u^{(j+1)}_{c-} & -u^{(j+1)}_{d-} & -u^{(j+1)}_{p-} & -u^{(j+1)}_{q-} \\
 \p u^{(j)}_{c+} & \p u^{(j)}_{d+} & \p u^{(j)}_{p+} & \p u^{(j)}_{q+} & -\p u^{(j+1)}_{c+} & -\p u^{(j+1)}_{d+} & -\p u^{(j+1)}_{p+} & -\p u^{(j+1)}_{q+} \\
 \p u^{(j)}_{c-} & \p u^{(j)}_{d-} & \p u^{(j)}_{p-} & \p u^{(j)}_{q-} & -\p u^{(j+1)}_{c-} & -\p u^{(j+1)}_{d-} & -\p u^{(j+1)}_{p-} & -\p u^{(j+1)}_{q-}  
\end{pmatrix} \cdot 
\begin{pmatrix}
c^{(j)}_\delta \\
d^{(j)}_\delta \\
p^{(j)}_\delta \\
q^{(j)}_\delta \\
c^{(j+1)}_\delta \\
d^{(j+1)}_\delta \\
p^{(j+1)}_\delta \\
q^{(j+1)}_\delta 
\end{pmatrix} \right|_{r=r_j} = 0.
\label{dnsol}
    \end{align}
\end{widetext}
where $j = 1,2,\cdots, n$. It is evident that the system contains $4n+2$ independent variables but only $4n$ linear equations. Therefore, two additional constraints from the ingoing modes are required to uniquely determine the solution, which is precisely what is expected. Note that Eq.\eqref{dnsol} is valid for the case with general ingoing modes.

\begin{figure}
    \centering
    \includegraphics[height=10.2cm]{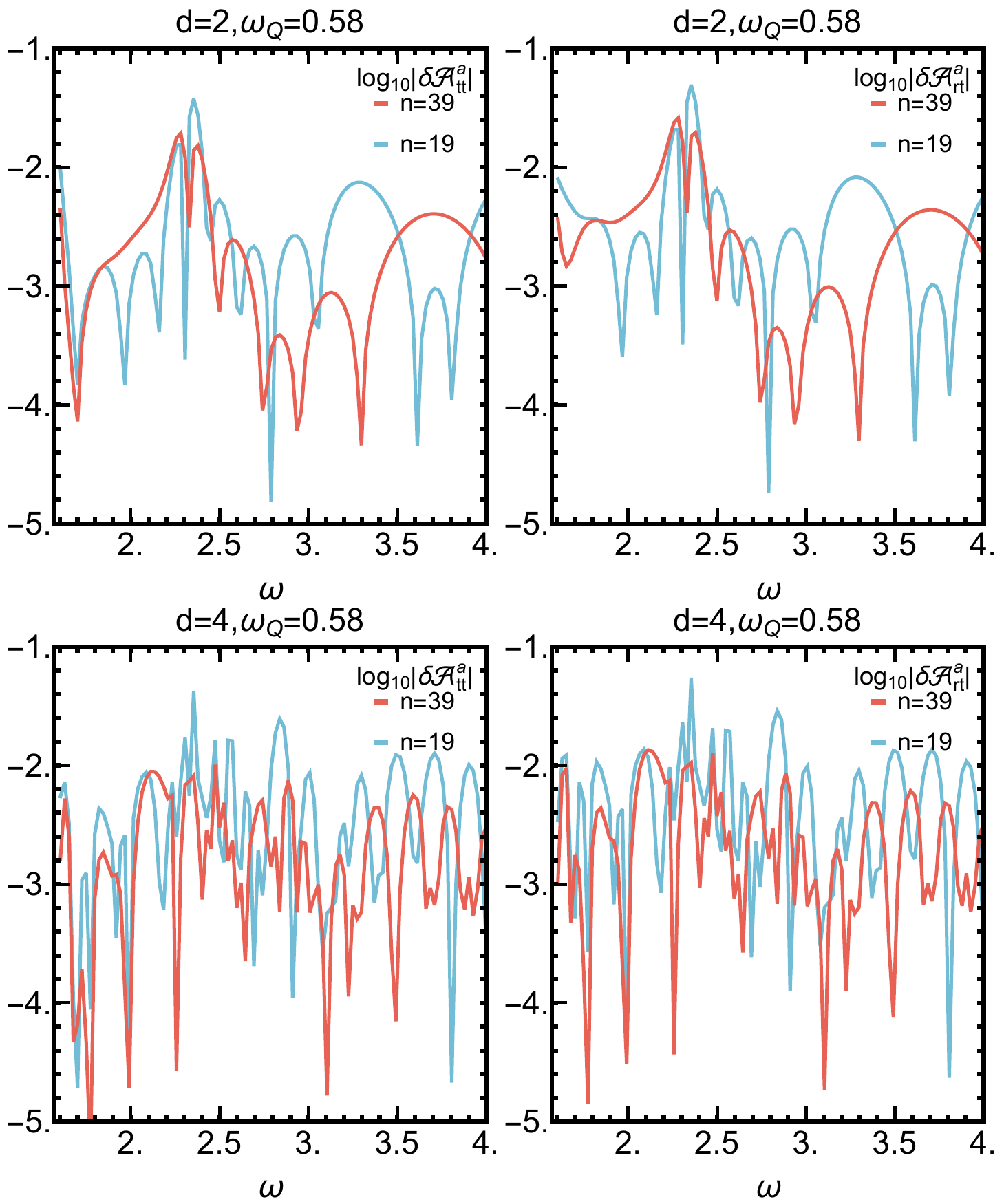}
    \caption{
    Differences between the analytical and numerical results for amplification factors with $g=1/3$. The analytical results are obtained by solving Eqs.~\eqref{dnsol} for a $(n+1)$-step background field. The numerical solutions are treated as reference (true) values. 
    }
 \label{anydn}
\end{figure}

In general, the analytical solution becomes lengthy for a large $n$, not particularly illuminating except for certain limits. Nevertheless, in practical terms, this analytical approach can notably improve numerical efficiency. For example, obtaining 100 exact numerical results takes about 400 seconds, while computing 500 results with $n=39$ using the current approach requires only about 40 seconds. This efficiency makes it easier to evaluate the amplification factors and to gain insight into the system, should a large parameter survey is required.

In Fig.~\ref{anydn}, we present a few comparisons between the analytical and the purely numerical results for the amplification factors. The analytical solutions are obtained from Eqs.~\eqref{dnsol}, where the background configuration is approximated by the $(n+1)$-step function. As $n$ increases, the step-function approximation for a general $Q$-ball systematically approaches the exact results, thereby validating our method.

Finally, note that in the case of single ingoing mode, Eq.~\eqref{nout} remains valid, if one replaces the $\cos\thi_H$'s with the corresponding quantities at $r_*\to r_n$. This again allows us to connect $\mc{A}^a$ with $\mc{A}^b$ via Eq.~\eqref{imp1}. In doing this, note that the matching condition at $r_*=r_n$ for the general $Q$-ball case contains $(c_{0}^{(n)}, d_{0}^{(n)}, p_{0}^{(n)}, q_{0}^{(n)})$, rather than $(c_{0}, d_{0})$. However, $(p_{0}^{(n)}, q_{0}^{(n)})$ can always be expressed as linear combinations of $(c_{0}^{(n)}, d_{0}^{(n)})$, with the corresponding coefficients fixed by the previous matching conditions at $r_j$ for $j = 1,2,\ldots,n-1$.

\section{Conclusion}
\label{sec:level6}

We have analytically investigated the superradiant amplification of waves scattered by a $Q$-ball. Our analysis sheds light on the previously unexplained multi-peak structure in the amplification spectra, which becomes most transparent in the case of large $Q$-balls.

Our analysis relies on approximating the background $Q$-ball with a discrete multi-step function, and is applicable to both thick-wall and thin-wall $Q$-balls in various spacetime dimensions. The perturbation solution is obtained by recognizing that the series expansion can be resummed as a linear transformation of Bessel functions. The scattering scenario is then imposed by matching to the asymptotic scattering waves. For thick-wall $Q$-balls, our method, with a semi-analytical treatment in the final step, provides a significant speed-up in evaluating the amplification factors. Comparisons between the present analytical and previous purely numerical results are presented, confirming the consistency and reliability of our analytical approach.

The analytical method is most tractable when applied to a single ingoing wave scattering off a large, thin-wall $Q$-ball. In this scenario, the amplification factors can be expressed in terms of simple sinusoidal functions, which explains the origin of the multi-peak structure in the spectra. Also taking the large $\oi$ limit, it is found that the amplification factor's dependence with $\oi$ reduces to a single damped sinusoidal function, whose frequency is determined by the $Q$-ball size, on top of a base $1/\oi^3$ term (see \eref{centralRes2}),
\bal
\mc{A} &\sim 1+  \f{S_2}{\oi^3} +  \f{S_1}{\oi^4} \cdot \cos [\text{(Q-ball size)} \cdot \omega +\text{(phase)} ].
\nonumber
\eal
Therefore, larger $Q$-balls thus exhibit more peaks in the amplification spectrum. Moreover, the $S_i$ above is modulated by the background $Q$-ball's frequency $\oi_Q$ and is proportional to $\sin [\text{(Q-ball size)} \cdot \omega_Q+\text{(phase)}]^i$.

We have also examined the close relation between the amplification factors and reflection rates. It is found that the extrema of these two are largely aligned with each other, which also originates from the sinusoidal nature of the reflection rates and from the fact that superradiant amplification arises from mode conversion between the two types of waves involved in the scattering. Moreover, our analytical solution enables us to determine the precise physical upper bounds on the amplification factors.

\acknowledgments

We would like to thank Qi-Xin Xie for helpful discussions. SYZ acknowledges support from the National Natural Science Foundation of China under grant No.~12475074, No.~12075233 and No.~12247103.


\appendix

\section{Amplification factor vs outgoing particle number}
\label{sec:levela}

In this appendix, we clarify the relation between the energy amplification factor and the outgoing particle number. 

In Fig.~\ref{exac}, the extrema of the amplification factor and the outgoing particle number are seen to almost coincide. However, they do not coincide exactly, but exhibit nontrivial relations. To see this, let us first focus on Case b, where $N^{in}_- = 1$ and $N^{in}_+ = 0$.

In this case, the energy amplification factor $\mathcal{A}_{tt}^b$ given in Eq.~\eqref{attb} can be expressed as a linear combination of $N_-^{out}$ and $E_+/E_-$. For convenience, we redefine the quantities as:
\begin{align}
    p(\omega)=\frac{E_+}{E_-}.
\end{align}
With this definition, the energy amplification factor takes the form,
\begin{align}
    \mathcal{A}_{tt}^b(\omega)=p(\omega)+(1-p(\omega))N_-^{out}(\omega),
\end{align}
where $p(\omega)>0$ and $N_-^{out}(\omega)\in[0,1]$, as implied by Eq.~\eqref{cons3}. The function $p(\omega)$ increases on the interval $(1+\omega_Q, \omega_0] \cup [\omega_0,\omega_1)$ and decreases on $(\omega_1,\infty)$, where $\omega_0$ denotes the point at which $p(\omega)=1$ and $\omega_1$ denotes its turning point. These characteristic values are explicitly given by:
\begin{align}
    \omega_0 & = \sqrt{1 + \omega_Q^2 + \sqrt{1 + 4\omega_Q^2}}, \\
    \omega_1 & = \sqrt{\frac{5}{2} + \omega_Q^2 + \frac{1}{2}\sqrt{17+32\omega_Q^2}}. 
\end{align}
Note that in the decreasing region $(\omega_1,\infty)$, the function $p(\omega)$ remains strictly greater than $1$. To investigate the oscillatory behavior of the energy amplification factor, we calculate the derivative of $\mathcal{A}_{tt}^b$ with respect to $\omega$:
\begin{align}
    (\mathcal{A}_{tt}^b(\omega))'=(1-p(\omega))(N_-^{out}(\omega))'+(1-N_-^{out}(\omega))p'(\omega).
\end{align}
Throughout this appendix, and only here, the prime denotes the derivative with respect to $\omega$.

The extrema of the energy amplification factor occur at the turning point $\omega_s$, which are determined by
\begin{align}
    (N_-^{out}(\omega_s))'=\frac{1-N_-^{out}(\omega_s)}{p(\omega_s)-1}p'(\omega_s).
    \label{osci:relat}
\end{align}
The extrema of the outgoing particle number, on the other hand, are located at points $\omega_t$, which satisfy $(N_-^{out}(\omega_t))'=0$. Since $N_-^{out}(\omega)\in[0,1]$, any point with $N^{out}_-(\omega)=1$ necessarily corresponds to the extrema of the outgoing particle number. 

\begin{table}[ht]
    \centering
    \caption{
    Relation between amplification factor ($\omega_s$) and particle number ($\omega_t$) peaks.
    }
    \begin{tabular}{ c c c c c }
    \hline \hline
         Region & $p(\omega_s),p'(\omega_s)$ & $(N_-^{out}(\omega_s))'$ & \text{max}$(\mathcal{A}_{tt}^b)$ & \text{min}$(\mathcal{A}_{tt}^b)$ \\
         \hline
          \multirow{2}{*}{$(1+\omega_Q,\omega_0)$} & $1>p(\omega_s)>0$ & \multirow{2}{*}{$(N_-^{out})'\leq0$} & ~\multirow{2}{*}{$\omega_s \leq \omega_t$}~ & ~\multirow{2}{*}{$\omega_s \geq \omega_t$}~ \\
          & $p'(\omega_s)>0$ &  &  &  \\
          \multirow{2}{*}{$(\omega_0,\omega_1)$} & $p(\omega_s)>1$ & \multirow{2}{*}{$(N_-^{out})' \geq 0$} & ~\multirow{2}{*}{$\omega_s \geq \omega_t$}~ & ~\multirow{2}{*}{$\omega_s \leq \omega_t$}~ \\
          & $p'(\omega_s)>0$ &  &  &  \\
          \multirow{2}{*}{$(\omega_1,\infty)$} & $p(\omega_s)>1$ & \multirow{2}{*}{$(N_-^{out})' \leq 0$} & ~\multirow{2}{*}{$\omega_s \leq \omega_t$}~ & ~\multirow{2}{*}{$\omega_s \geq \omega_t$}~ \\
          & $p'(\omega_t)<0$ &  &  &  \\
          \hline \hline
          Region & $p(\omega_s),p'(\omega_t)$ & $(N_+^{out}(\omega_s))'$ & \text{max}$(\mathcal{A}_{tt}^a)$ & \text{min}$(\mathcal{A}_{tt}^a)$ \\
         \hline
          \multirow{2}{*}{$(1+\omega_Q,\omega_0)$} & $p(\omega_s)>1$ & \multirow{2}{*}{$(N_+^{out})' \leq 0$} & ~\multirow{2}{*}{$\omega_s \geq \omega_t$}~ & ~\multirow{2}{*}{$\omega_s \leq \omega_t$}~ \\
          & $p'(\omega_s)<0$ &  &  &  \\
          \multirow{2}{*}{$(\omega_0,\omega_1)$} & $1>p(\omega_s)>0$ & \multirow{2}{*}{$(N_+^{out})' \geq 0$} & ~\multirow{2}{*}{$\omega_s \leq \omega_t$}~ & ~\multirow{2}{*}{$\omega_s \geq \omega_t$}~ \\
          & $p'(\omega_s)<0$ &  &  &  \\
          \multirow{2}{*}{$(\omega_1,\infty)$} & $1>p(\omega_s)>0$ & \multirow{2}{*}{$(N_+^{out})' \leq 0$} & ~\multirow{2}{*}{$\omega_s \geq \omega_t$}~ & ~\multirow{2}{*}{$\omega_s \leq \omega_t$}~ \\
          & $p'(\omega_t)>0$ &  &  &  \\
          \hline \hline
    \end{tabular}
    \label{tab:appenb}
\end{table}

Let us tabulate the relations between $\omega_s$ and $\omega_t$ for the three intervals $(1+\omega_Q,\omega_0)$, $(\omega_0,\omega_1)$, and $(\omega_1,\infty)$ separately. In Table~\ref{tab:appenb}, we summarize the relation between $\omega_a$ and $\omega_b$ for different cases. Here, ${\rm max}(\mathcal{A}_{tt}^b)$ and ${\rm min}(\mathcal{A}_{tt}^b)$ denote, respectively, the local maxima and minima of the energy amplification factor for Case b. In addition, results for Case a are also included in the table. 
For the energy flux amplification factors, analogous results can be obtained. By redefining $p(\omega)=\omega_+/(-\omega_-)$ and analyzing Eq.~\eqref{osci:relat}, one arrives at conclusions similar to those for the energy amplification factor, which we do not elaborate on here.

\section{ Case of $d=1$ }
\label{sec:levelb}

This appendix addresses the case $d=1$, which is an exception to the general formulas presented in the main text in the sense that it allows for a new branch of solutions with odd parity. That is, due to the absence of the first-derivative term in the radial perturbative equation of motion, the perturbation field can have a nonzero first derivative as  $r \to 0$, allowing for solutions with odd parity. For even-parity solutions, where the first derivative vanishes at the origin, the general method introduced in the main body can be extended to the $d=1$ case.

In this case, the perturbative equations~\eqref{EOM::pte} simplify to
\begin{align}
    \p_r^2 \eta_\pm + (k_\pm^2-U)\eta_\pm-W\eta_\mp^*=0, \notag
\end{align}
As $r \to \infty$, the asymptotic behavior of the solution is 
\begin{align}
    \eta_\pm(\omega,r\rightarrow\infty)\rightarrow A_\pm e^{ik_\pm r} + B_\pm e^{-ik_\pm r}, \notag
\end{align}
where $A_\pm$ and $B_\pm$ are constants. Using the approximate profile~\eqref{app::pro} and applying the power series expansion near the origin, the absence of the regularity conditions~\eqref{asymp1} means that the coefficients $c_1^{(1)}$ and $d_1^{(1)}$ can be non-zero. Therefore, the perturbative scattering solutions can be decomposed into components of different spatial parities: odd-parity and even-parity. For even parity, as $r\to-\infty$, we have
\begin{align}
    A_+=B_+ \text{ and } A_- = B_-,
\end{align}
and in region $(1)$, among the four independent parameters $c_0^{(1)},d_0^{(1)},c_1^{(1)}$, and $d_1^{(1)}$, only $c_0^{(1)}$ and $d_0^{(1)}$ are non-zero. For odd parity, as $r\to-\infty$, 
\begin{align}
    A_+=-B_+ \text{ and } A_- = - B_-,
\end{align}
and in region $(1)$, only $c_1^{(1)}$ and $d_1^{(1)}$ are non-zero. The recurrence relations reduce to
\begin{align}
\!
\begin{pmatrix}
c_{l+2}^{(j)} \\
d_{l+2}^{(j)}
\end{pmatrix} = \frac{1}{(l+2)(l+1)} \begin{pmatrix}
U^{(j)}-k_+^2  & W^{(j)} \\
W^{(j)}  & U^{(j)}-k_-^2
\end{pmatrix}
\begin{pmatrix}
c_{l}^{(j)} \\
d_{l}^{(j)}
\end{pmatrix}. \!
\end{align}
This structure resembles the expansion of trigonometric functions, and hence, the solutions take the form (for $r\ge0$)
\begin{align}
& \begin{pmatrix}
\eta_+^{(j)} \\
(\eta_-^{(j)})^*
\end{pmatrix} = 
\lambda^{-1} 
\begin{pmatrix}
\cos (\sqrt{-\rho_1} r)  & 0 \\
0  & \cos (\sqrt{-\rho_2} r)
\end{pmatrix} \lambda 
\begin{pmatrix}
c_{0}^{(j)} \\
d_{0}^{(j)}
\end{pmatrix} \notag \\
& \quad + \lambda^{-1} 
\begin{pmatrix}
\frac{\sin (\sqrt{-\rho_1} r)}{\sqrt{-\rho_1}}  & 0 \\
0  & \frac{\sin (\sqrt{-\rho_2} r)}{\sqrt{-\rho_2}}
\end{pmatrix} \lambda
\begin{pmatrix}
c_{1}^{(j)} \\
d_{1}^{(j)}
\end{pmatrix},
\label{dnsoldim1}
\end{align}
where $j=1,2,\cdots,n$, and $\rho_1,\rho_2$, and $\lambda$ depend on $W^{(j)}$ and $U^{(j)}$. For $j=1$, according to the analysis in Section~\ref{sec:level3::2}, the solution involves only two independent initial parameters: $c_0^{(1)}$ and $d_0^{(1)}$ for even parity, or $c_1^{(1)}$ and $d_1^{(1)}$ for odd parity. For $j \ge2$, the coefficients $c_0^{(j)},d_0^{(j)},c_1^{(j)}$, and $d_1^{(j)}$, together with $A_\pm$ and $B_\pm$, are determined by the matching conditions.

\begin{figure}
    \centering
    \includegraphics[height=5.1cm]{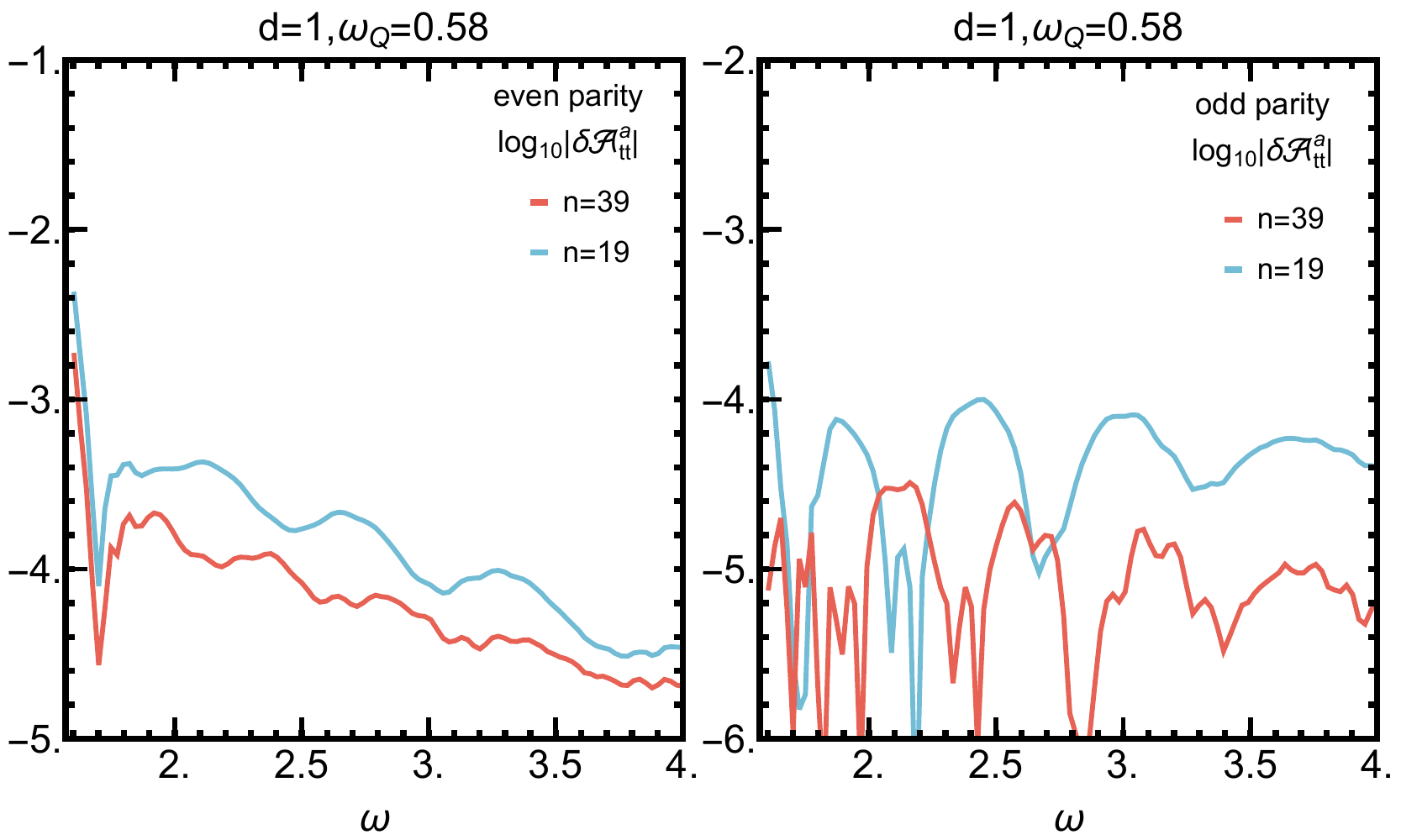}
    \caption{
    Differences between the analytical and numerical results for amplification factors with $d=1$ and $g=1/3$ for $(n+1)$-step $Q$-balls. 
    }
 \label{anydndim1}
\end{figure}

For $n\ge2$ and $r\ge0$, in a region $(j)$ with $j\ge2$, the recurrence relation Eq.~\eqref{rec::1} reduces the solution to a linear combination of the four parameters $c_0^{(j)},d_0^{(j)},c_1^{(j)}$, and $d_1^{(j)}$:
\begin{align}
    \! & \eta^{(j)}_+(r) = c_0^{(j)} u^{(j)}_1 + d_0^{(j)} u^{(j)}_2 + c_1^{(j)} u^{(j)}_3 + d_1^{(j)} u^{(j)}_4, \!  \\
    & \! \left( \eta^{(j)}_-(r) \right)^* = c_0^{(j)} u^{(j)}_5 + d_0^{(j)} u^{(j)}_6 + c_1^{(j)} u^{(j)}_7 + d_1^{(j)} u^{(j)}_8, \! \notag
\end{align}
where $u^{(j)}_\ell$ with $\ell=1,2,\cdots,8$ are all functions of $W^{(j)},U^{(j)}$, and $r$. These four parameters are solely determined by the matching conditions. Considering the parity transformation $r \to -r$, one obtains the following constraints: for even parity,
\begin{align}
    \eta^{(j)}_+(-r) = \eta^{(j)}_+(|r|), ~ \eta^{(j)}_-(-r) = \eta^{(j)}_-(|r|),
\end{align}
while for odd parity, 
\begin{align}
    \eta^{(j)}_+(-r) = -\eta^{(j)}_+(|r|), ~ \eta^{(j)}_-(-r) = -\eta^{(j)}_-(|r|).
\end{align}
Therefore, the perturbative solution of the case $d=1$ is established.

In Fig.~\ref{anydndim1}, we present the difference between the analytical and numerical results for amplification factors for the two parity modes. The results show that our analytical solutions exhibit excellent agreement with the purely numerical ones.


\bibliographystyle{apsrev4-1}
\bibliography{zref2}

\end{document}